\documentstyle[epsf]{mn}

\title[Field Ellipticals at $z\approx 0.3$ III]{The properties of
field elliptical galaxies at intermediate redshift.  III: the
Fundamental Plane and the evolution of stellar populations from
$z\approx0.4$ to $z=0$.}

\author[T.~Treu et al.]{T.~Treu,$^{1,2,4}$ M.~Stiavelli,$^{2}$
G.~Bertin,$^1$ S.~Casertano$^{2}$ and P.~M{\o}ller$^{3}$\\
$^1$ Scuola Normale Superiore, P.za dei Cavalieri 7, I-56126, Pisa,
Italy\\
$^2$ Space Telescope Science Institute, 3700 San Martin Dr.,
Baltimore, MD 21218, USA\\
$^3$ ESO, Karl-Schwarzschild Str. 2, D85748, Garching bei M\"unchen,
Germany\\
$^4$ present address: California Institute of Technology, Astronomy 105-24, Pasadena, CA 91125, USA; e-mail tt@astro.caltech.edu\\ 
}

\newcommand{\tx}[1]{\textrm{#1}}
\newcommand{\pc}[1]{\protect\citename{#1}}

\newcommand{\kms}{km~$\tx{s}^{-1}$}

\newcommand{\Io}{I$_{814}$}
\newcommand{\Vs}{V$_{606}$}
\newcommand{\sbe}{$SB_{\tx{e}}$}
\newcommand{\resec}{$r_{\tx{e}}$}
\newcommand{\Rekpc}{$R_{\tx{e}}$}

\begin{document}
\maketitle

\begin{abstract}

We report on the study of a sample of 25 {\it field} early-type
galaxies, in the redshift range $z\approx0.1-0.5$, selected on the
basis of colours and morphology from the Medium Deep Survey (MDS) of
the Hubble Space Telescope (HST). Surface photometry in two colors
(F606W and F814W) and redshifts have been derived for all the galaxies
in the sample, while velocity dispersions has been measured for 19 of
the sample galaxies, as described in a companion paper.

Our study of the evolution of the Fundamental Plane (FP) with redshift
yields the following results.  Field early-type galaxies define a
tight FP out to $z\approx0.4$, with scatter unchanged with respect to
local samples, within the observational errors. The intermediate
redshift FP is offset with respect to the local FP of the Coma
Cluster, in the sense that, for given effective radius and velocity
dispersion, galaxies are brighter than expected from the local
relation. The offset of the FP is found to increase with redshift. The
range of parameters covered by our sample is not sufficiently extended
to measure the slopes of the FP at intermediate redshift. Similar
results are found for the \sbe-\Rekpc\, relation, out to
$z\approx0.5$.

The evolution of the FP (and of the \sbe-\Rekpc\, relation) is studied
quantitatively with a Bayesian-Montecarlo technique.  By applying this
technique, we find that the offset of the intercept of the FP ($\Delta
\gamma$) with respect to the local FP increases as $\Delta \gamma =
\tau z$ with the following 68 per cent limits: $0.33<\tau<0.44$ (for
$\Omega=1, \Omega_{\Lambda}=0$) or $0.44<\tau<0.56$ (for $\Omega=0.3,
\Omega_{\Lambda}=0.7$).  In addition, we interpret the results in
terms of the evolution of the stellar populations, under the
assumption of passive evolution. In a single-burst scenario, the
observed properties are consistent with those of a stellar population
formed at $z\ga2$ (for $\Omega=1, \Omega_{\Lambda}=0$, $H_0=50$ \kms
Mpc$^{-1}$) or $0.8<z<1.6$ (for $\Omega=0.3, \Omega_{\Lambda}=0.7$,
$H_0=65$ \kms Mpc$^{-1}$).  If a small fraction of the stellar mass is
formed in a secondary burst, the primary burst may have occurred at
higher $z$. For example (for $\Omega=0.3, \Omega_{\Lambda}=0.7$,
$H_0=65$ \kms Mpc$^{-1}$) the primary burst may have occurred at
$z\ga3$ if a secondary burst with a tenth of stellar mass occurred at
$0.6<z<0.8$.

Finally, the intercept and scatter of the FP found for field
early-type galaxies and for cluster data (taken from the literature)
at $z\approx0.3-0.4$ are mutually consistent, within the observational
errors. If higher redshift (up to $z=0.83$) cluster data are
considered, the ages of the stellar populations of field early-type
galaxies inferred from a single-burst scenario are found to be
marginally smaller than the ages derived for the cluster galaxies.

\end{abstract}
\begin{keywords}

galaxies: elliptical and lenticular, cD---galaxies:
evolution---galaxies: photometry---galaxies:
kinematics and dynamics---galaxies: fundamental
parameters---galaxies: formation

\end{keywords}

\section{Introduction}
\label{sec:intro}

Despite the progress made in understanding the physics of early-type
galaxies (E/S0), a single widely accepted model, capable of explaining
all their observed properties, is not yet available. A variety of
scenarios are still considered to be viable (see e.g. Larson 1975,
Toomre 1977, van Albada 1982). Schematically, we can identify the
following two main classes of scenarios. In the traditional view, the
monolithic collapse scenario, E/S0 form in a rapid collapse shortly
after an intense burst of star formation. Since the stellar
populations of E/S0 in the local Universe are generally old, and no
such intense bursts of star formation are observed at redshift $z<1$,
it is generally assumed that most of these monolithic collapse events
occurred at high redshifts.  In the alternate view, the merging
scenario, E/S0 form by interactions of disc galaxies. The merging
scenario has been cast into a cold dark matter cosmological framework,
the hierarchical clustering scenario (White \& Rees 1978; Kauffman,
White \& Guiderdoni 1993).  In the hierarchical clustering scenario
the formation of massive E/S0 by merging of discs is a relatively
recent phenomenon (Kauffman 1996).

In the absence of a satisfactory theoretical scenario, new
observations play a key role, by providing additional input and the
opportunity to test predictions. Since the theoretical models have
been conceived and `calibrated' in order to reproduce the properties
of local galaxies, the observation of intermediate ($0.1<z<1$) and
high redshift ($z>1$) galaxies is particularly important.

Several observational results have been obtained on this subject in
recent years. Most of them have been obtained for the cluster
environment, where early-type galaxies are more abundant and studied
more efficiently (Ellis et al.\, 1997; Dressler et al. 1997; Stanford,
Eisenhardt \& Dickinson, 1998; van Dokkum et al.\ 1998a,b; Brown et
al.\ 2000; Kelson et al.\ 2000b; van Dokkum et al.\ 2000). These
studies indicate that the stellar populations of early-type galaxies
in the core of rich clusters have possibly formed at high redshift,
earlier than $z\sim 2-3$.

When comparing intermediate redshift E/S0 to local ones, one should
take into account the evolution with redshift of clusters and of the
population of galaxies in clusters.  Although the evolution of
clusters of galaxies is still somewhat controversial (Postman et al.\
1996), it is is likely that the population of galaxies in clusters
evolves by encroaching on isolated galaxies and small groups. Thus,
part of the progenitors of present-day cluster E/S0 were likely
located outside clusters in the past. Therefore, to gain a reliable
global picture, it is necessary to study the evolution of E/S0 in all
environments, from the core of rich clusters to the field. In
addition, E/S0 have been observed to be subject to interactions with
their environment ranging from weak distant encounters to major
merging events. Thus, since the frequency of interactions depends upon
the environment, evolution should depend upon the environment as well,
if interactions play a significant role. For this reason, the role of
environment should be especially important in the merging scenario. In
the hierarchical clustering scenario, numerical simulations predict
(Kauffmann 1996) that field E/S0 should have younger stars than
cluster ones and have assembled later.  Therefore, the comparison of
the evolution of the properties of cluster and field E/S0 provides a
test of a prediction of hierarchical clustering models.

So far, only a few studies of field early-type galaxies, based on
HST high resolution images, have been published (Im et al.\ 1996;
Treu \& Stiavelli 1999; Schade et al.\ 1999; Menanteau et al.\ 1999;
Brinchmann \& Ellis 2000), often with very limited statistics
(e.g. the analysis of early-type galaxies in the Hubble Deep Fields,
Franceschini et al.\ 1998; Benitez et al.\ 1999; Kodama, Bower \&
Bell, 1999; see also Treu et al.\ 1999). One important result is that
field early-type galaxies at intermediate redshifts span a wider range
of colours than cluster ones (more extended to the blue; Schade et
al.\ 1999). This is likely to result from a more complex evolution of
the stellar populations, with episodes of star formation occurring at
lower redshifts than in clusters (see e.g. Trager et al. 2000a,b). A
detailed study of the star formation history of field E/S0 is needed
to clarify this issue.

A correct picture of the evolution of stellar populations is also
needed in order to understand how the number density of E/S0 evolves
with redshift. In fact, several studies (Treu \& Stiavelli 1999;
Menanteau et al.\ 1999) find that the number of red morphologically
selected E/S0 at high $z$ is smaller than what is predicted by models
based on passive luminosity evolution and constant comoving density
models. However, Schade et al.\ (1999) find that the number density of
early-type galaxies, selected purely on morphology without color
criteria, is consistent with being constant from $z\sim1$ to
$z\sim0.2$ (see also Im et al.\ 1996). Similarly, Menanteau et al.\
(1999) find that, if all E/S0 are considered (including the blue
ones), the data are consistent with no evolution of the number
density.

The evolutionary history of E/S0 derived from the data depends
critically on the detailed star formation history involved (for other
sources of uncertainties see, e.g., Chiba \& Yoshii 1999; Schade et
al.\ 1999; Barger et al.\ 1999; Daddi et al.\ 2000).  In fact, even
otherwise negligible amounts of young stars are sufficient to make
high$-z$ E/S0 appear blue and thus to produce the observed ``deficit''
of red high-z E/S0 (Treu \& Stiavelli 1999; see also Jimenez et al.\
1999 and Abraham et al.\ 1999).

More detailed spectroscopic information is needed to study the
evolution with redshift of the structural properties of E/S0 and of
their stellar populations.  For this reason we have initiated (Treu et
al.\ 1999; hereafter T99) an observational project aimed at obtaining
high signal-to-noise medium resolution spectra for a sample of field
early-type galaxies at intermediate redshift selected on the basis of
HST-WFPC2 images. The photometric and kinematic measurements for the
entire sample ($0.1\la z\la 0.6$) are described in a companion paper
(Treu et al.\ 2001; hereafter PII).  In the present paper, the third
of the series, we study the evolution with redshift of the Fundamental
Plane (Dressler et al.\ 1987; Djorgovski and Davis 1987), a very tight
empirical correlation between the main observables of E/S0
(Section~\ref{sec:FP}). Its existence and tightness have profound
implications in terms of galactic formation and evolution, and it is
therefore particularly interesting to investigate how far in the past
it exists and what is the time evolution of its coefficients and
scatter. In addition, the Fundamental Plane is used as a diagnostic of
stellar populations, to study the star formation history of field
E/S0. A similar study is performed on the evolution of the
\sbe-\Rekpc\, relation (Kormendy 1977; Section~\ref{ssec:HK}), a
photometric projection of the Fundamental Plane.  Future papers will
report on the measurement of ages and metallicity, using the
absorption line indices in the Lick/IDS system (Trager et al.\ 1998).

The paper is organized as follows.  The Fundamental Plane as a
diagnostic tool of the stellar populations is described in Section
\ref{sec:FP}. In Section~\ref{sec:data} we give a brief summary of the
data described in PII.  In Section \ref{sec:obs} we illustrate the
observational results, i.e. we discuss the existence and location of
the intermediate redshift field Fundamental Plane.  In
Section~\ref{sec:sel} we analyse selection effects that must be taken
into account while interpreting the results in terms of stellar
population evolution (Section \ref{sec:evo}). In Section \ref{sec:clu}
we compare the Fundamental Plane of field early-type galaxies to the
Fundamental Plane of clusters at similar redshift, with data and
results taken from the literature. Conclusions are drawn in Section
\ref{sec:consum}

The results presented here are given for two sets of values of the
cosmological parameters (the matter density $\Omega$ and the
cosmological constant $\Omega_{\Lambda}$, expressed in dimensionless form,
and the Hubble constant H$_0=50h_{50}$\kms Mpc$^{-1}$): a long lived
$\Lambda$ Universe ($\Omega=0.3$, $\Omega_{\Lambda}=0.7$,
$h_{50}$=1.30; age of the Universe $\sim14.6$ Gyr) and the `classical
choice' of parameters ($\Omega=1$, $\Omega_{\Lambda}=0$, $h_{50}=1$;
age of the Universe $\sim 13.1$ Gyr). We will refer to them as the
$\Lambda$ cosmology and the classical cosmology respectively.

\section{The Fundamental Plane of early-type galaxies}

\label{sec:FP}

Early-type galaxies are a rather homogeneous family of galaxies. It has
been early-on recognized that several observables, describing their
dynamical state, size, and chemical composition, correlate with
luminosity. Typical examples are the effective radius ($R_{\tx{e}}$;
Fish 1964), the central velocity dispersion ($\sigma$; Faber \&
Jackson 1976), the effective surface brightness (\sbe; Binggeli,
Sandage \& Tarenghi 1984), the integrated colours (Bower, Lucey \&
Ellis 1992 and references therein), and metal indices such as Mg$_2$
(Bender, Burstein \& Faber 1993 and references therein).

The relatively large scatter of these correlations suggests that a
`second parameter' could be involved. The `second parameter' was
found when large spectrophotometric surveys during the mid-eighties
revealed the existence of a tight relation between effective radius,
effective surface brightness, and central velocity dispersion
(Djorgovski \& Davis 1987; Dressler et al.\ 1987).  The relation,
\begin{equation}
\label{eq:FP} 
\log R_{\tx{e}} = \alpha \log~\sigma + \beta~SB_{\tx{e}} + \gamma,  
\end{equation} 
where $R_{\tx{e}}$ is in kpc (we will reserve the notation \resec\,
when such radius is measured in arcsec), $\sigma$ in \kms, and
$SB_{\tx{e}}$ in $\tx{mag arcsec}^{-2}$, was called the Fundamental
Plane (FP). Because the value of the Hubble constant enters the
relation through the calculation of the effective radius in kpc, the
value of $\gamma$ depends on $H_0$. In the following we will call
FP-space the three-dimensional space defined by $\log$\Rekpc,
$\log\sigma$, and \sbe.

The quantities \sbe\, and \Rekpc\, are found to depend on the
wavelength of observation, as a result of colours and colour gradients
in the stellar populations of early-type galaxies. When needed, we
will explicitly state the dependence on wavelength by adding a
subscript (e.g., $R_{\tx{e}B}$ is the effective radius in the B band).
Not much is known about systematic variations of $\sigma$ with the
spectral range used for the measurements. In PII we showed that no
significant variation is noticed if lines such as Ca H and K or NaD
are included or excluded from the spectral region of the
measurement. In the following, in order to compare our data to data
from the literature, we will assume that the value of the central
velocity dispersion does not depend on the spectral region used for
the measurement.

The intercept $\gamma$ depends on wavelength. For example, in the Coma
Cluster, Bender et al.\ (1998) report in the B band
$\gamma_{B}=-8.895$ assuming $h_{50}=1$; in the V band, using the
sample and slopes of Lucey et al.\ (1991), $\gamma_{V}=-8.71$ is
found. The determination of $\gamma$ is hampered by the errors in the
determination of distances and extinction. No differences, within the
observational errors, have been found among the intercepts (and
slopes) of the FP of nearby clusters and the field (Pahre, de Carvalho
\& Djorgovski 1998). 

The slope $\beta$ is very well determined, independently of the
different selection criteria and fitting techniques adopted by various
groups; in addition, it does not change significantly with wavelength
(Pahre, Djorgovski \& de Carvalho 1998; Scodeggio et al.\ 1998). A
typical value is $\beta=0.32$. The slope $\alpha$ suffers from larger
uncertainties and it is observed to vary with wavelength (Pahre et
al.\ 1998b ; Scodeggio et al.\ 1998).  Accurate measurements over
large samples of galaxies have shown that the observed scatter of the
FP is larger than the observational errors (e.g. J{\o}rgensen, Franx
\& Kj{\ae}rgaard 1996). In other words, a significant fraction ($\sim
0.07-0.08$ in $\log$ \Rekpc) of the observed scatter is due to scatter
in the physical properties of E/S0. Within the observational errors,
the intrinsic scatter is observed to be constant with wavelength and
along the FP.

\subsection{The physical origin of the FP}

\label{sec:physFP}

The very existence of the FP is a remarkable fact. Any theory of
galaxy formation and evolution must be able to account for its
tightness and its universality. In the following we
will address the issues of how the observables entering the FP may be
related to intrinsic physical quantities such as mass and
luminosity of early-type galaxies and we will list some of the
physical processes that have been proposed to be at the basis of the
FP. In addition we will describe how the existence of the FP and its
tightness can be used to derive information on the formation and
evolution of early-type galaxies.
 
Let us define an effective mass $M$,
\begin{equation}
M\equiv \sigma^2 R_{\tx{e}}/G,
\label{eq:masdef}
\end{equation} 
where $G$ is the gravitational constant. If homology holds,
i.e. early-type galaxies are structurally similar, the total mass
${\mathcal M}$ (including dark matter if present) is proportional to
$M$
\begin{equation}
{\mathcal M}=c_1 M
\label{eq:dynhol}
\end{equation}
and we can interpret Eq.~\ref{eq:masdef} in terms of the Virial
Theorem. If $c_1$ depends on mass, then we will say that weak homology
holds.

Let us also define the effective luminosity $L$ as
\begin{equation}
-2.5\log L  = SB_{\tx{e}}-5\log R_{\tx{e}}-2.5\log 2\pi.
\label{eq:homology}
\end{equation}
In general, from the FP and Equations~\ref{eq:masdef} and
\ref{eq:homology} we have:
\begin{equation}
L \propto 10^{\frac{\gamma}{2.5\beta}}  M^{\frac{\alpha}{5\beta}} 
R_{\tx{e}}^{\frac{10\beta-\alpha-2}{5\beta}}.
\label{eq:MvsL}
\end{equation}
The observed values for $10\beta-\alpha-2$ cluster around to zero,
although there is significant scatter that could be related to
differences in the selection of the samples, in the determination of
distances, in the fitting techniques, and extinction corrections. If
\begin{equation}
10\beta-\alpha-2=0, 
\label{eq:slope}
\end{equation}
then Equation~\ref{eq:MvsL} can be reduced to a power law
relation between effective mass and luminosity (e.g. Faber et al.\
1987; van Albada, Bertin \& Stiavelli 1995):
\begin{equation}
L \propto M^{\eta},
\label{eq:pl}
\end{equation}
where
\begin{equation}
\eta=0.2\frac{\alpha}{\beta}.
\end{equation}
Based on these definitions, the $M/L$ (effective mass-to-light ratio)
of a galaxy is readily obtained in terms of the FP observables:
\begin{equation}
M/L \propto 10^{0.4 SB_{\tx{e}}} \sigma^2 R_{\tx{e}}^{-1}.
\label{eq:ML}
\end{equation}
By comparing $M/L$ to the value predicted by the FP for the same
effective radius and velocity dispersion,
\begin{equation}
(M/L)_{FP} \propto 10^{-\frac{\gamma}{2.5\beta}}
\sigma^{\frac{10\beta-2\alpha}{5\beta}}R_{\tx{e}}^{\frac{2-5\beta}{5\beta}},
\label{eq:MLFP}
\end{equation}
the scatter of the FP can be connected to the scatter in $M/L$ of the
early-type galaxy population:
\begin{equation}
\Sigma_i \left[ \log (M/L^i) - \log (M/L)_{FP}^i \right]^2 = \frac{\Sigma_i \left(\gamma^i - \gamma
\right)^2}{(2.5 \beta)^2}
\label{eq:scatFP}
\end{equation}
where the superscript $i$ labels the galaxies of the sample under
consideration, and $\gamma^i$ is defined as
\begin{equation}
\gamma^i=\log R_{\tx{e}}^i - \alpha \log \sigma^i - \beta SB_{\tx{e}}^i.
\label{eq:gammai}
\end{equation}
Thus, the remarkable tightness of the FP implies a very
low scatter in the $M/L$ of early-type galaxies.

When a theoretical effort to understand this empirical scaling law is
made, it is necessary to model the relation between effective mass and
total mass (${\mathcal M}$) and the relation between stellar ($M_{*}$)
and total mass. For example, if we assume
\begin{equation}
{\mathcal M} \propto M \propto M_{*},
\label{eq:masses}
\end{equation}
then Equations \ref{eq:masdef} and \ref{eq:MLFP} yield
\begin{equation}
\frac{M_*}{L}\propto 10^{-\frac{\gamma}{2.5\beta}} {\mathcal
M}^{\frac{5\beta-\alpha}{5\beta}}
R_{\tx{e}}^{-\frac{10\beta-\alpha-2}{5\beta}}.
\label{eq:tilt}
\end{equation}
By neglecting the mild dependence on effective radius,
Equation~\ref{eq:tilt} implies that the stellar mass-to-light ratio
depends on the total mass of the galaxy. This phenomenon is sometimes
referred to as the tilt of the FP (Renzini \& Ciotti 1993; see also
Bender, Burstein \& Faber 1992; Pahre et al. 1998b).

Several interpretations in terms of properties of the stellar
populations have been suggested to explain the tilt of the FP (see,
e.g., Renzini \& Ciotti 1993; Ciotti, Lanzoni \& Renzini 1996; Graham
\& Colless 1997; Pahre et al.\ 1998b). However no conclusive
interpretation of the FP has been found so far. Some of the hypotheses
have turned out to be insufficient to explain the FP (e.g. the tilt
cannot be attributed entirely to a relation between metallicity and
mass, Pahre et al.\ 1998b). On the other hand, the viable hypotheses
require an `unnatural' fine tuning in order to explain the tilt while
preserving the tightness.

The remarkable tightness of the FP can be used to set constraints on
the formation and evolution of early-type galaxies. For example,
assuming weak homology and a proportionality between total mass and
stellar mass, we know from the FP that at any given mass the scatter
in $\log(M_*/L)$ is very small ($\sim$0.10). If we attribute such
scatter entirely to scatter in age, we can set an upper limit to it
($\Delta t / t$). Similarly, we can set an upper limit to the scatter
in metallicity ($\Delta Z / Z$). Unfortunately, correlations between
the various effects can conspire to keep the scatter small, so that,
for example, a small scatter in $\log(M_*/L)$ does not necessarely
imply a small scatter in both age and metallicity.

The absence of a difference between the FP of field and cluster E/S0
is also a remarkable fact. For example, within the scheme described in
this section, it can be interpreted in terms of the absence of
significant differences between the average $M/L$, and thus between
the stellar populations, of field and cluster E/S0. This fact is
consistent with the very small difference found between the
Mg-$\sigma$ relation (in which the observables are distance
independent) of a large sample of local ellipticals in the cluster and
in the field environments by Bernardi et al.\ (1998).

A very effective way to further investigate the origin of the FP, and
thus to learn about the evolution of E/S0, is to study the FP as a
function of redshift. In fact, with this information it is possible to
distinguish between models that predict an evolution of the slopes
with redshift, e.g. the model where the tilt is due to Initial Mass
Function (IMF) effects (see Renzini \& Ciotti 1993), and models that
do not, e.g. the model where the tilt is due to dark matter
distribution. Similarly for the scatter: if the scatter is due to
scatter in the ages of the stellar populations, then it should
increase with look-back time; if it is due to scatter in metallicity,
it should not.

\subsection{The FP at intermediate redshift as a diagnostic of the evolution 
of stellar populations}

\label{sec:FPz}

Recently it has been observed that early-type galaxies in clusters at
intermediate redshift also populate a tight FP (van Dokkum \& Franx
1996; Kelson et al.\ 1997; van Dokkum et al.\ 1998b; Bender et al.\
1998; Pahre 1998; J{\o}rgensen et al.\ 1999; Kelson et al.\ 2000b).

The intercept of the FP ($\gamma$) is observed to vary as a function
of redshift. So far, no dramatic change in the slopes with redshift
has been reported, even though an accurate measurement is difficult,
because of the small samples available and of the intrinsic problems
in this measurement, such as completeness corrections and fitting
technique (see, e.g., van Dokkum \& Franx 1996; T99; J{\o}rgensen et
al.\ 1999, Kelson et al.\ 2000b). For example, Kelson et al.\ (2000b),
using the largest sample at intermediate redshift ($z=0.33$) published
at the moment of this writing (30 E/S0 galaxies in a single cluster),
find $\alpha=1.31\pm0.13$ and $\beta=0.344\pm0.10$ in the V band;
these values are indistinguishable from those found locally. The
absence of a dramatic change in the slopes argues against the
interpretation of the FP solely in terms of an age-mass relation.

A simple explanation of these findings can be given by recalling the
correspondence of the FP observables to $M$ and $M/L$.  Let us
consider a sample of galaxies with observables labelled by superscript
$i$. Let us assume that effective radius and central velocity
dispersion do not change on the observed time-scale.  For each galaxy,
we can compute the variation of the logarithm of the effective
mass-to-light ratio with redshift,
\begin{equation}
\Delta \log(M/L)^i\equiv \log(M/L)_{z=z_2}^i-\log(M/L)_{z=z_1}^i,
\label{eq:DMLz}
\end{equation}
in terms of the variation of combinations of the coefficients of the
FP with $z$ from Eq.~\ref{eq:MLFP}:
\begin{eqnarray}
\label{eq:DMLzpar}
\Delta \log(M/L)^i=  & \Delta \left(\frac{10\beta-2\alpha}{5\beta}\right) \log \sigma^i \nonumber\\&  + \Delta \left(\frac{2-5\beta}{5\beta}\right)\log \tx{R}_{\tx{e}}- \Delta \left(\frac{\gamma^i}{2.5\beta}\right),
\end{eqnarray}
where the symbol $\Delta$ indicates the difference of the quantity at
two redshifts as in Equation~\ref{eq:DMLz}. 

For the analysis presented here we will assume that $\alpha$ and
$\beta$ are constant. This assumption is broadly consistent with the
observations and makes the interpretation of the results
straightforward. In fact, if $\alpha$ and $\beta$ are constant, we
have that:
\begin{equation}
\Delta \log \left( M/L^i \right)=-\frac{\Delta \gamma^i}{2.5\beta},
\label{eq:MLzg}
\end{equation}
i.e. the evolution of $\log (M/L^i)$ depends only on the evolution of
$\gamma^i$.  In practice, by measuring $\gamma^i$ for a sample of
$N_g$ galaxies at intermediate redshift, and by comparing it to the
value of the intercept found in the local Universe ($\gamma_0$), we
can measure the average evolution of $\log (M/L)$ with cosmic time:
\begin{equation}
\label{eq:aML}
<\Delta \log (M/L)>=-\Sigma_{i=1}^{i=N_g}
\frac{\gamma^i-\gamma_0}{2.5\beta N_g},
\end{equation}
i.e., if the intercept $\gamma$ is computed as the average of $\gamma^i$, 
\begin{equation}
\label{eq:aML2}
<\Delta \log (M/L)>=-\frac{\Delta \gamma}{2.5\beta}.
\end{equation}
Similarly, the intrinsic scatter of $M/L$ is related to observable
quantities by Equation~\ref{eq:scatFP}, and therefore its evolution
with redshift can be measured. Note that, by assuming constant
\Rekpc\, and $\sigma$, we are neglecting any structural and dynamical 
evolution of the galaxies; in addition, by assuming constant $\alpha$
and $\beta$, we obtain that the evolution of $M/L$ is independent of
$M$.

In the spirit of this observational study, we will focus our attention
on the measurement of the evolution of the effective mass-to-light
ratio $M/L$ and we will not attempt to address the issues associated
with the relation between the effective mass $M$ and the total mass
${\mathcal M}$. In particular, we will limit our study to the
evolution of stellar populations, neglecting any dynamical or
structural evolution. In other words, the only properties of the
galaxies that will be allowed to vary in our study will be those of
the integrated stellar population. The comparison with the
expectations of passive evolution models will be done by assuming
that:
\begin{equation}
M\propto M_*.
\label{eq:MMstar}
\end{equation}

Two effects must be kept in mind when applying this diagnostic:

\begin{enumerate}
\item {\bf Morphological evolution}. If galaxy types evolve
into one another, then local early-type galaxies are not simply the
evolution of intermediate redshift ones. For example, if spirals
evolve into early-type galaxies as suggested by hierarchical
clustering scenarios, then the FP at intermediate redshift can look
tight and slowly evolving because E/S0 are being recognized only when
their stellar populations are already old.
\item {\bf Cluster population evolution}. The population of galaxies
within clusters is likely to evolve with redshift. If this happens,
then the comparison of cluster and field samples can be biased.

\end{enumerate}

A complete analysis of these effects requires a comprehensive
theoretical model, able to describe the joint evolution of galaxies
and of their environment. Such modeling is beyond the aims of this
observational study. Therefore, we will mostly limit our presentation
to the observational results (to be used as a constraint when testing
models). In order to facilitate comparison with theoretical models,
the sample selection process has been defined as clearly as possible
in PII. Some considerations on the population biases in the context of
simple passive evolution of the stellar populations are presented in
Section~\ref{ssec:aeo}.

The comparison of the stellar populations in cluster and field
environment is particularly interesting.  In fact, as mentioned in the
Introduction, one of the expectations of hierarchical clustering
models is that early-type galaxies and their stellar populations form
much later in the field than in the core of rich
clusters. Unfortunately, in the local Universe, in addition to the
observational uncertainties listed in Section~\ref{sec:FP}, the
detection of a possible difference between cluster and field galaxies
with the FP is hampered by the generally old age of their stellar
populations.  It is easier to detect any possible difference at
intermediate redshift, where the FP is in principle more sensitive. In
fact, since peculiar velocities (the motion with respect to the local
comoving system) are limited (a few hundreds of \kms, see e.~g. Groth,
Juszkiewicz \& Ostriker 1989), the uncertainty on distance
determination becomes negligibly small with increasing redshift. In
addition, if the stellar populations of field and cluster E/S0 had
different average ages, any difference in the FP should increase with
redshift thus becoming detectable.

\subsection{The \sbe-\Rekpc\, relation}

\label{sec:proj}

The correlation associated with the projection of the FP on the
photometric plane, also known as the \sbe-\Rekpc\, relation (Kormendy
1977),
\begin{equation}
SB_{\tx{e}} =a_1\log R_{\tx{e}}+a_2
\label{eq:HKdef}
\end{equation}  
is very useful for our purposes because it does not involve
internal kinematics. Therefore, it can be followed to higher redshifts
and fainter absolute magnitudes. The drawback is that the
\sbe-\Rekpc\, relation has a larger intrinsic scatter than the FP and
its coefficients are much more sensitive to sample selection biases
(see, e.g., Capaccioli, Caon \& D'Onofrio 1992).  In the following we
will use the evolution of the \sbe-\Rekpc\, relation to extend the
findings achieved with the FP to higher redshift and to fainter
absolute luminosities.

\section{The data}
\label{sec:data}

The galaxies have been selected on the basis of HST-WFPC2 images taken
from the HST-Medium Deep Survey (Griffiths et al.\ 1994), mainly based
on colour, magnitude, and morphology. The sample selection criteria
are extensively discussed in PII and are taken into account in the
analysis presented here. Briefly, we can say that the sample analyzed
in this paper is representative of the population of early-type
galaxies at $z\approx0.1-0.5$ with $0.95<$~\Vs-\Io~$<1.7$ and
\Io~$<19.3$ (as in PII \Vs\, and \Io\, indicate Vega magnitudes
through HST filters F606W and F814W respectively).

The sample is a field sample in the sense that the galaxies are
selected on random WFPC2 pointings, excluding the regions of known
clusters. On the basis of the images available, it is not possible to
identify and to exclude members of groups or galaxies in the outskirts
of rich clusters. In this respect, our operational definition of field
does not coincide with the definition used in theoretical
investigations, where the full three dimensional and dynamical
information available is used to identify rich clusters, groups, and
field (e.g., Kauffmann 1996).

The data, the data reduction, and the caveats of the photometric and
spectroscopic measurements are described in PII. For quick reference,
we summarise in Table~\ref{tab:data} the relevant observed quantities
for the sample of E/S0 analyzed in this paper.

\begin{table*}
\caption{Summary of observed quantities. For each galaxy (gal) we list
redshift, central velocity dispersion ($\sigma$, if measured; in
\kms), effective radius (\resec; in arcsec), and effective surface
brightness ($SB_{\tx{e}}$; in mag arcsec$^{-2}$) in the rest frame B
and V bands. Details on the measurements are given in PII. Note that
the errors on the effective radius and the effective surface
brightness are correlated: the combination that enters the FP,
$r_{\tx{e}}-\beta SB_{\tx{e}}$, is better determined than the
individual quantities (for a discussion see PII; see also
Figures~\ref{fig:FPso1} and \ref{fig:FPso0.3}).}
\label{tab:data}
\begin{tabular}{ccccccc}
gal	& z	& $\sigma$ & $r_{\tx{eB}}$ & $SB_{\tx{eB}}$ & $r_{\tx{eV}}$ & $SB_{\tx{eV}}$ \\
\hline
A1 & 0.147 & 242$\pm$24 & 1.59$\pm$0.14 & 20.54$\pm$0.08 & 1.64$\pm$0.07& 19.67$\pm$0.13  \\
D1 & 0.385 & 309$\pm$28 & 0.78$\pm$0.10 & 19.80$\pm$0.20 & 0.70$\pm$0.07& 18.77$\pm$0.16  \\
E1 & 0.294 & 211$\pm$28 & 0.89$\pm$0.09 & 20.78$\pm$0.15 & 0.93$\pm$0.08& 19.91$\pm$0.21  \\
F1 & 0.295 &  -		& 1.11$\pm$0.00 & 21.53$\pm$0.07 & 1.05$\pm$0.00& 20.55$\pm$0.09  \\
G1 & 0.295 & 234$\pm$22 & 1.59$\pm$0.22 & 20.78$\pm$0.20 & 1.51$\pm$0.12& 19.80$\pm$0.18  \\
I1 & 0.293 & 202$\pm$23 & 0.64$\pm$0.12 & 19.96$\pm$0.26 & 0.64$\pm$0.07& 19.06$\pm$0.24  \\
C2 & 0.425 &    -       & 1.04$\pm$0.10 & 20.21$\pm$0.16 & 1.00$\pm$0.09& 19.42$\pm$0.12  \\
D2 & 0.192 &    -       & 0.37$\pm$0.32 & 18.68$\pm$0.96 & 0.40$\pm$0.17& 18.08$\pm$0.74  \\
E2 & 0.398 & 233$\pm$16 & 1.61$\pm$0.08 & 20.78$\pm$0.09 & 1.48$\pm$0.08& 19.77$\pm$0.10  \\
F2 & 0.336 & 252$\pm$23 & 1.00$\pm$0.09 & 20.73$\pm$0.14 & 0.94$\pm$0.05& 19.74$\pm$0.11  \\
A3 & 0.231 & 201$\pm$18 & 1.08$\pm$0.13 & 20.60$\pm$0.14 & 0.97$\pm$0.07& 19.42$\pm$0.14  \\
B3 & 0.490 &  -         & 0.23$\pm$0.01 & 17.57$\pm$0.18 & 0.23$\pm$0.03& 16.98$\pm$0.31  \\
C3 & 0.106 & 155$\pm$19 & 0.48$\pm$0.08 & 19.74$\pm$0.14 & 0.47$\pm$0.05& 18.78$\pm$0.10  \\
E3 & 0.263 &  -         & 1.31$\pm$0.16 & 21.70$\pm$0.12 & 1.30$\pm$0.10& 20.87$\pm$0.16  \\
G3 & 0.408 & 259$\pm$17 & 0.81$\pm$0.16 & 20.17$\pm$0.36 & 0.77$\pm$0.17& 19.19$\pm$0.40  \\
I3 & 0.410 & 207$\pm$16 & 0.62$\pm$0.11 & 19.74$\pm$0.28 & 0.65$\pm$0.10& 18.93$\pm$0.23  \\
M3 & 0.337 & 243$\pm$19 & 0.90$\pm$0.13 & 20.23$\pm$0.18 & 0.83$\pm$0.08& 19.25$\pm$0.13  \\
R3 & 0.348 & 243$\pm$15 & 1.20$\pm$0.09 & 20.64$\pm$0.12 & 1.16$\pm$0.11& 19.71$\pm$0.21  \\
S3 & 0.356 & 230$\pm$23 & 1.16$\pm$0.11 & 21.22$\pm$0.29 & 0.86$\pm$0.05& 19.60$\pm$0.07  \\
T3 & 0.117 & 142$\pm$30 & 1.18$\pm$0.14 & 21.05$\pm$0.20 & 1.27$\pm$0.08& 20.29$\pm$0.08  \\
A4 & 0.117 & 178$\pm$19 & 0.99$\pm$0.20 & 20.85$\pm$0.02 & 0.95$\pm$0.11& 19.75$\pm$0.13  \\
D4 & 0.285 & 238$\pm$21 & 1.14$\pm$0.06 & 20.67$\pm$0.05 & 1.11$\pm$0.04& 19.73$\pm$0.07  \\
H4 & 0.340 & 231$\pm$16 & 1.65$\pm$0.12 & 21.17$\pm$0.10 & 1.58$\pm$0.06& 20.25$\pm$0.10  \\
I4 & 0.337 & 258$\pm$27 & 0.65$\pm$0.17 & 20.26$\pm$0.38 & 0.65$\pm$0.09& 19.43$\pm$0.24  \\
B4 & 0.211 &      -   	& 0.54$\pm$0.03 & 20.43$\pm$0.06 & 0.62$\pm$0.02& 20.06$\pm$0.03  \\
\hline			      
\end{tabular}		      
\end{table*}

\section{Observational results}

\label{sec:obs}

In this section we will discuss the location of our sample of field
early-type galaxies at intermediate redshift in the FP-space in
comparison to the location of local samples. Only galaxies that obey
the selection criteria listed in PII are
considered.

In principle, one would like to compare the intermediate redshift data
to field local data. However, as discussed in Section~\ref{sec:FP}, no
significant difference has been found so far in the local Universe
between the FP in the field and that in the cluster environment,
within the accuracy allowed by the uncertainties in the determination
of distances, reddening corrections, and the small size of the
available samples \cite{PdCD98}. For these reasons, we will compare
the intermediate redshift data to the FP of the Coma cluster, a well
defined relation, with selection biases under control.

The discussion will be extended to a larger sample of galaxies in
Section~\ref{ssec:HK} by means of the projection of the FP on the
photometric plane, the \sbe-\Rekpc\, relation, including galaxies
without measured velocity dispersion.  

\subsection{The Fundamental Plane of field early-type galaxies at intermediate redshift}

\label{ssec:FPf}

\begin{figure}
\mbox{\epsfysize=8cm \epsfbox{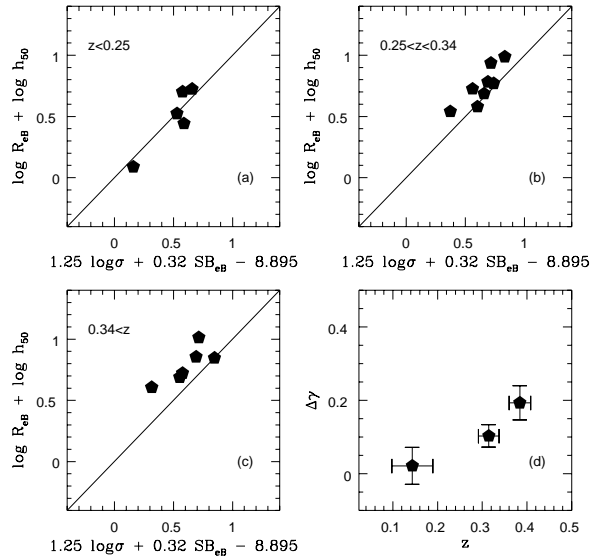}}
\caption{Fundamental Plane in the rest-frame B band at intermediate
redshift. In panels (a), (b), and (c) we plot the intermediate
redshift galaxies (pentagons) binned in redshift. The solid line
represents the FP of Coma, as measured by Bender et al.\ (1998). In
panel (d) the average offset of the intercept $\gamma$ (at fixed
slopes $\alpha$ and $\beta$, taken from Bender et al.\ 1998) is
plotted as a function of redshift. The error bars are the standard
deviation of the mean. The classical cosmology is assumed.}
\label{fig:FPBo1}
\end{figure}

\begin{figure}
\mbox{\epsfysize=8cm \epsfbox{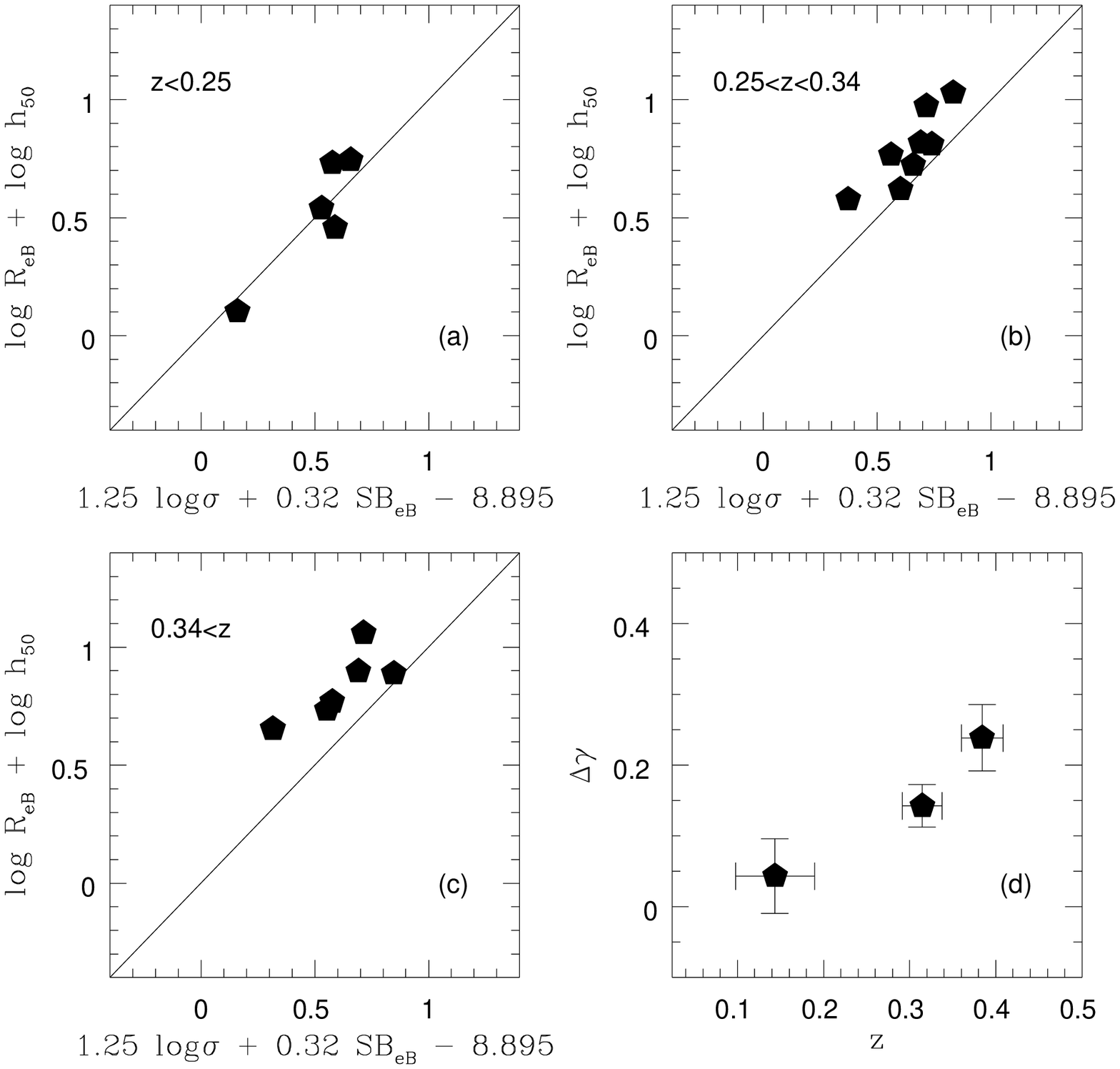}}
\caption{As in Figure~\ref{fig:FPBo1}, in the $\Lambda$ cosmology.}
\label{fig:FPBo0.3}
\end{figure}

\begin{figure}
\mbox{\epsfysize=8cm \epsfbox{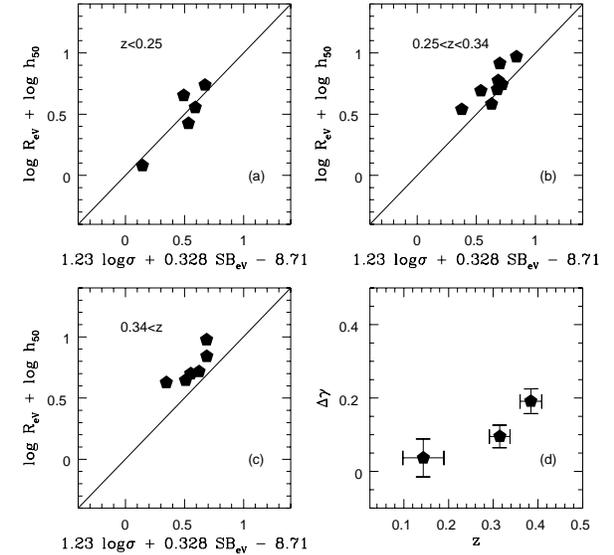}}
\caption{As in Figure~\ref{fig:FPBo1} (classical cosmology), data in
the rest-frame V band. The FP of Coma, drawn as a solid line, is taken
from Lucey et al.\ (1991).}
\label{fig:FPVo1}
\end{figure}

\begin{figure}
\mbox{\epsfysize=8cm \epsfbox{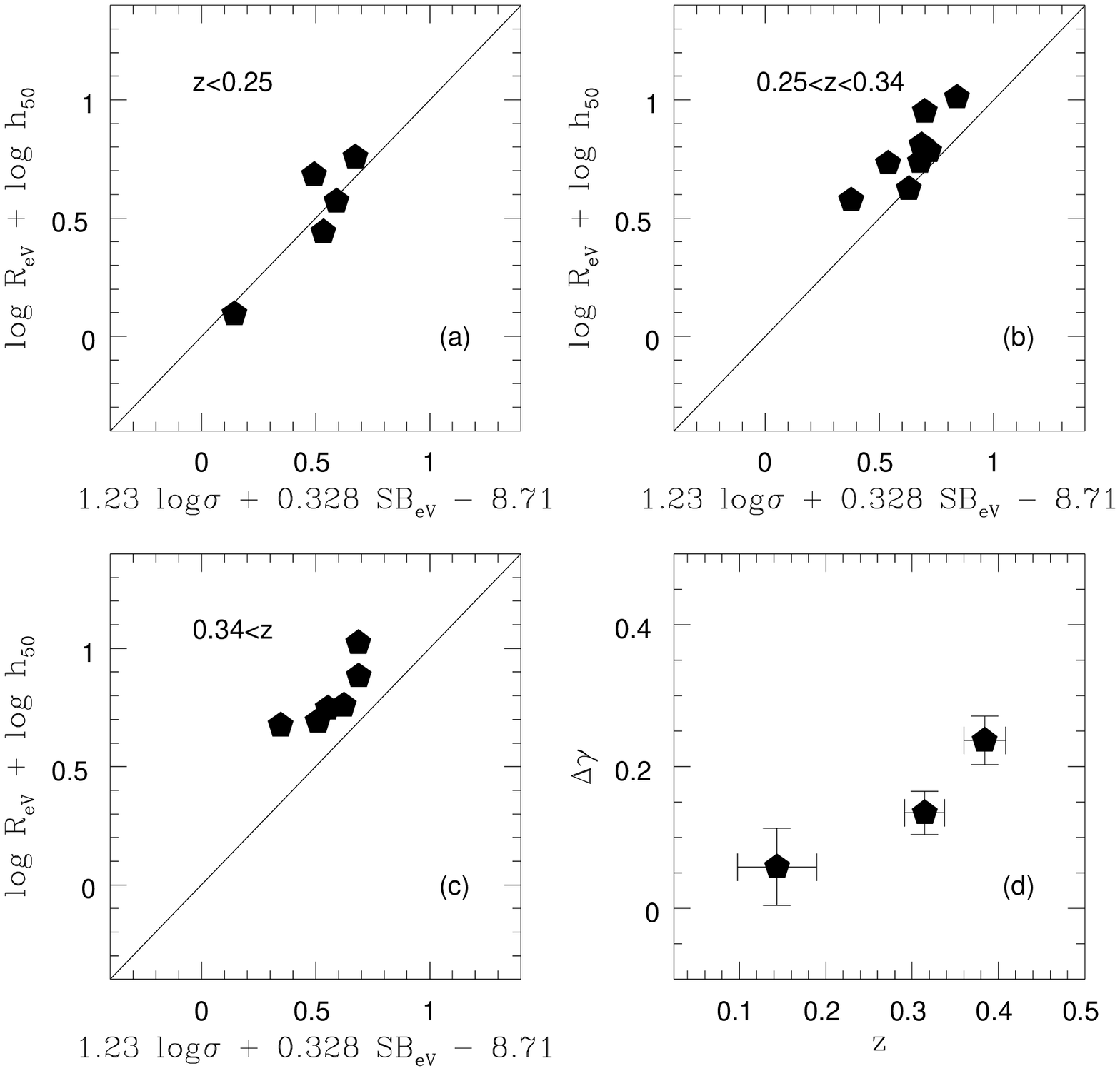}}
\caption{As in Figure~\ref{fig:FPVo1}, in the $\Lambda$ cosmology.}
\label{fig:FPVo0.3}
\end{figure}

\begin{figure}
\mbox{\epsfysize=8cm \epsfbox{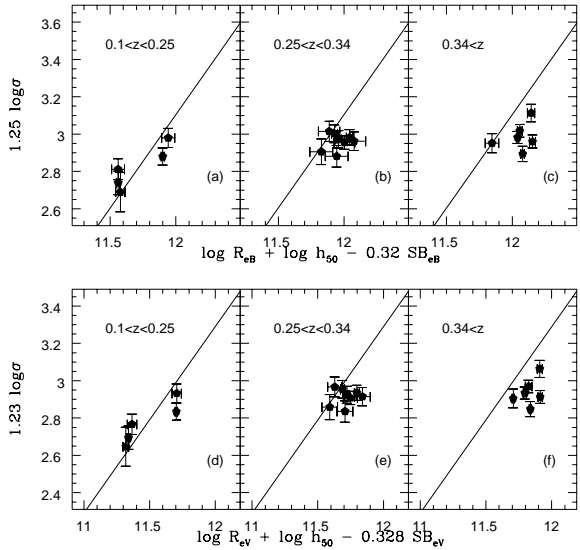}}
\caption{Evolution of the field FP in the B (upper panels) and V bands
(lower panels), in the classical cosmology. The symbols are as in
Fig~\ref{fig:FPBo1}. The Fundamental Plane is seen here in the
projection most natural for the description of errors, with
photometric and kinematic measurements on different axes.  The errors
on the photometric parameters are correlated; the combination that
enters the FP is particularly robust (see PII).}
\label{fig:FPso1}
\end{figure}

\begin{figure}
\mbox{\epsfysize=8cm \epsfbox{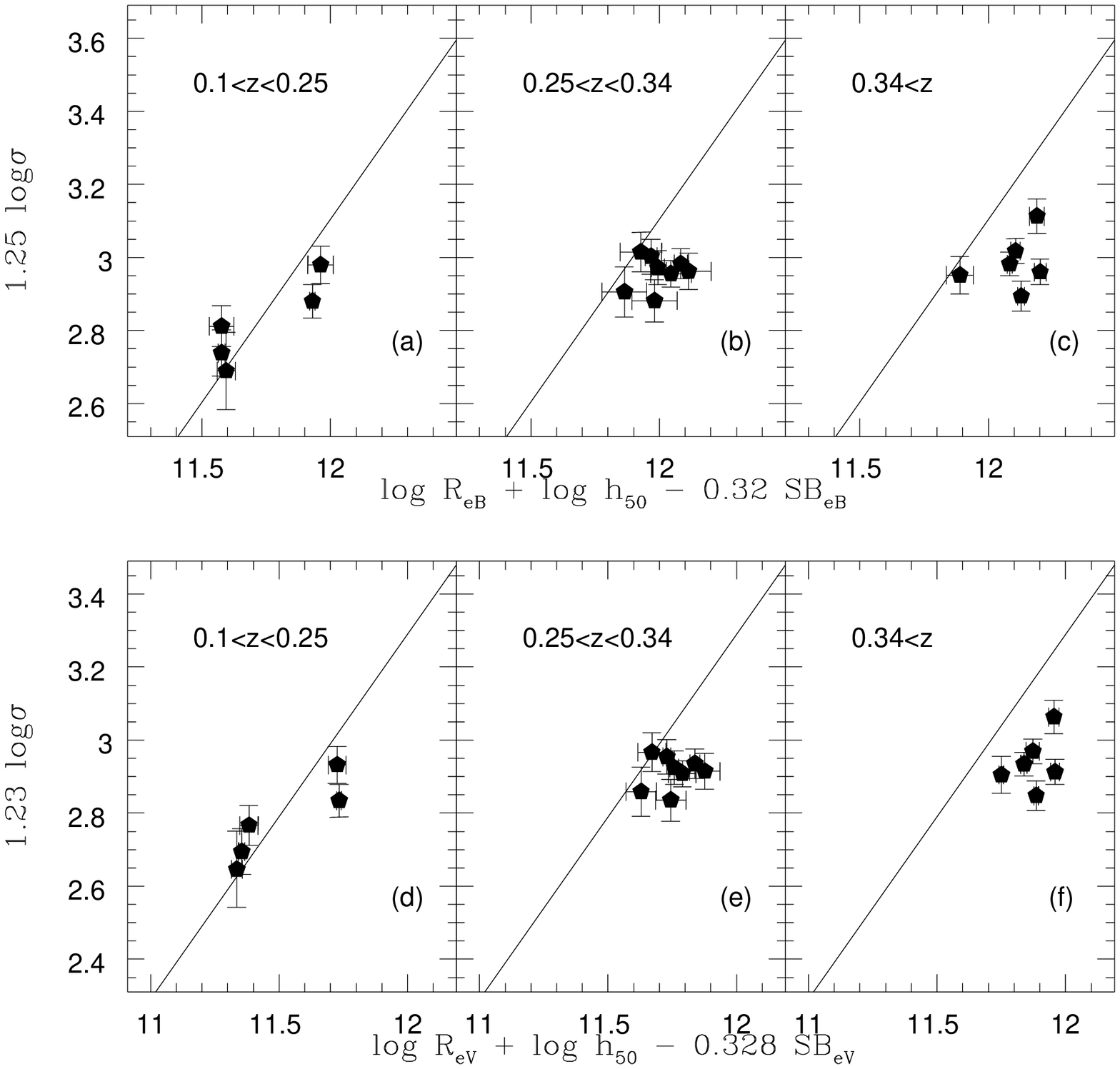}}
\caption{As in Figure~\ref{fig:FPso1}, in the $\Lambda$ cosmology.}
\label{fig:FPso0.3}
\end{figure}

In Figures~\ref{fig:FPBo1} to \ref{fig:FPVo0.3}, panels (a), (b), and
(c), we plot the location in the FP-space of the intermediate redshift
galaxies binned in redshift. The data are shown in the rest-frame B
and V bands and for two choices of values for the cosmological
parameters (see captions). Since errors on the photometric parameters
are correlated, we do not show them in this projection. A better
visualization of the errors is given in the projection that separates
kinematics and photometry, used in Figures~\ref{fig:FPso1} and
\ref{fig:FPso0.3}. In panels (d) of Figures~\ref{fig:FPBo1} to
\ref{fig:FPVo0.3} we show the evolution of the intercept $\gamma$ of
the FP with redshift.

Qualitatively, two main facts emerge:

\begin{enumerate}

\item {\bf Existence of the FP}. At any given redshift 
between $z=0$ and $z\approx0.4$ the FP of {\it field} early-type
galaxies is well defined. The scatter is small. No trend of increasing
scatter with redshift is noticeable within the accuracy allowed by the
small number statistics available.

\item {\bf Evolution of the FP}. At given \Rekpc\, and $\sigma$, the
intermediate redshift galaxies have brighter effective surface
brightness with respect to the values predicted by the local relation
(solid line) and hence are more luminous and have lower $M/L$. The
average offset increases with redshift, as shown in panels (d).

\end{enumerate}

To make these statements more quantitative, let us describe the
evolution of the intercept $\gamma$ (at fixed slopes $\alpha$ and
$\beta$) with a linear relation:

\begin{equation}
\Delta\gamma = \tau z.
\label{eq:gammev}
\end{equation}

\begin{figure}
\mbox{\epsfysize=8cm \epsfbox{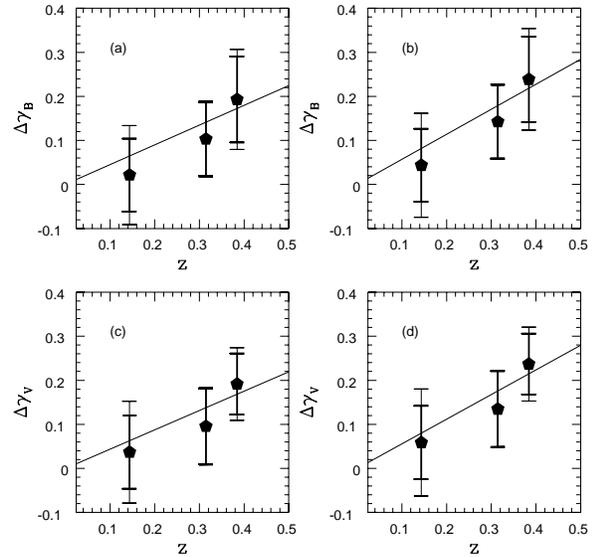}}
\caption{Evolution of the intercept $\gamma$ of the FP and its scatter with
redshift. The solid line is a least $\chi^2$ linear fit to the
data. The thin error bars are the measured scatter, the thick error
bars are the scatter corrected for its evolutionary component (see
Section~\ref{ssec:FPf}). The classical cosmology is used in panels (a)
and (c), the $\Lambda$ cosmology in panels (b) and (d).}
\label{fig:scatz}
\end{figure}

With a least $\chi^2$ fit (see Figure~\ref{fig:scatz}) applied to the
data in the V band we obtain $\tau=0.44\pm0.03$ (classical cosmology)
and $\tau=0.56\pm0.04$ ($\Lambda$ cosmology); in the B band we obtain
$\tau=0.45\pm0.04$ (classical cosmology) and $\tau=0.57\pm0.04$
($\Lambda$ cosmology). This description allows us to estimate the
scatter of the FP in any redshift bin. In fact, there are three
contributions to the scatter: the measurement error, the intrinsic
scatter, and the evolutionary drift. The last item is simply the
combined effect of the width of the redshift bins and the
evolution of the intercept. We can correct for the scatter induced by
the evolutionary drift by `evolving' the galaxies to the average
redshift of the bin with Equation~\ref{eq:gammev}. The corrected
scatter thus obtained is remarkably small and constant with redshift
(0.08-0.09 in $\gamma$). In Figure~\ref{fig:scatz} we plot the
evolution of the intercept $\gamma$ with redshift together with the
best-fitting line. The thin error bars represent the total scatter,
while the thick error bars represent the scatter corrected for
evolution.  It is harder to separate the contribution from measurement
errors, because they are correlated and include a systematic
component. However, by subtracting in quadrature the average
uncertainty on $\gamma$ (for a single galaxy this is 0.05-0.06) one
can estimate the intrinsic scatter to be approximately 0.05-0.07.
Given the limited number of galaxies available per redshift bin, the
error on the scatter is quite large ($\sim 30 $ per cent for a
Gaussian distribution, neglecting systematics). Nevertheless, a
substantial increase in the scatter can be ruled out. Larger samples
are needed in order to measure the scatter as a function of redshift
accurately.

\subsection{The \sbe-\Rekpc\, relation at intermediate redshift} 

\label{ssec:HK}

\begin{figure}
\mbox{\epsfysize=8cm \epsfbox{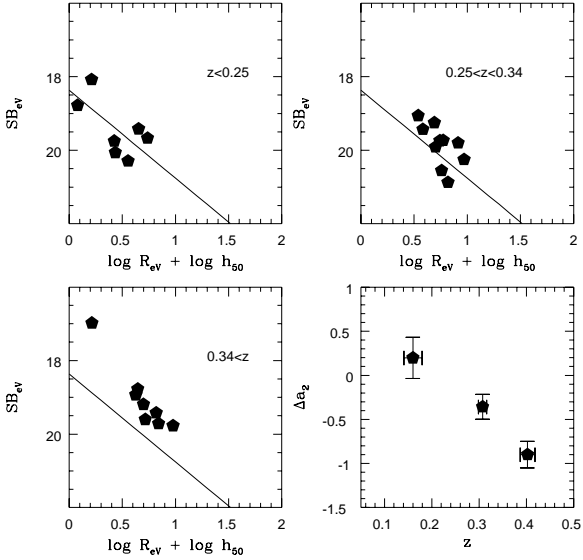}}
\caption{\sbe-\Rekpc\, relation for field early-type galaxies at
intermediate redshift in the V band. The solid line is the best fit to
the Coma data by Lucey et al.\ (1991). The classical cosmology has
been used to derive effective radii. The average offset is plotted in
panel (d) as a function of redshift. Error bars are the standard
deviation of the mean of the sample.}
\label{fig:KVo1}
\end{figure}

\begin{figure}
\mbox{\epsfysize=8cm \epsfbox{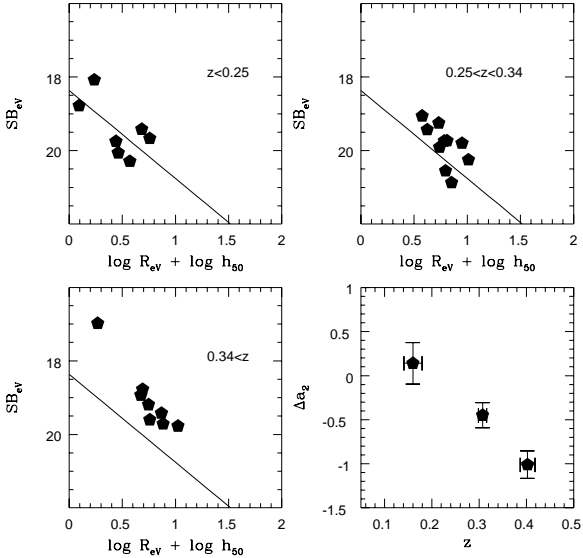}}
\caption{As in Figure~\ref{fig:KVo1}, in the $\Lambda$ cosmology.}
\label{fig:KVo0.3}
\end{figure}

We will now study the evolution of the \sbe-\Rekpc\, introduced in
Section~\ref{sec:proj} on the larger sample available without internal
kinematic information. The sample includes 25 galaxies out to
$z=0.49$.  The results in the V band are plotted in Figures~\ref{fig:KVo1}
and \ref{fig:KVo0.3}. The results in the B band are very similar (see
figures in Section~\ref{sec:evo}).

Following the analysis of the previous section, we describe the
evolution of the intercept of the \sbe-\Rekpc\, relation ($a_2$ of
Equation~\ref{eq:HKdef}) with a linear relation,
\begin{equation}
\Delta a_2=\tau ' z.
\label{eq:Kev}
\end{equation}
By means of a least $\chi^2$ fit, we find $\tau'=-1.30$ (classical
cosmology) and $\tau'=-1.59$ ($\Lambda$ cosmology) in the V band. In
the B band we find $-1.37$ and~$-1.67$ respectively.

In the context of passive evolution of stellar populations, with the
assumption that the stellar mass is proportional to the effective mass
$M_*\propto M$, Equations~\ref{eq:aML2} and~\ref{eq:gammev} yield:
\begin{equation}
<\Delta \log (M_*/L)> = - \frac{\tau z}{2.5 \beta}.
\label{eq:Mltau}
\end{equation}
Thus, if we assume that \Rekpc\, and $a_1$ (Equation~\ref{eq:HKdef})
are constant with redshift, and that the evolution of the
\sbe-\Rekpc\, relation is given by passive evolution of $M_*/L$, i.~e.
\begin{equation}
\Delta a_2 = \Delta SB_{\tx{e}} = 2.5 <\Delta \log (M_*/L)>,
\end{equation}
$\tau'$ should be related to $\tau$ by:
\begin{equation}
\tau = - \tau' \beta.
\end{equation}
The values found for $\tau$ and $\tau'$ are in excellent agreement
with this relation. Naturally, the agreement does not prove that
either result is correct, since the comparison depends on rather
strong assumptions. Nevertheless, this provides an obvious consistency
check for both the observations (that rely on independent local
relations) and the pure passive evolution interpretation.

\section{Selection effects}
\label{sec:sel}

\begin{figure}
\mbox{\epsfxsize=8cm \epsfbox{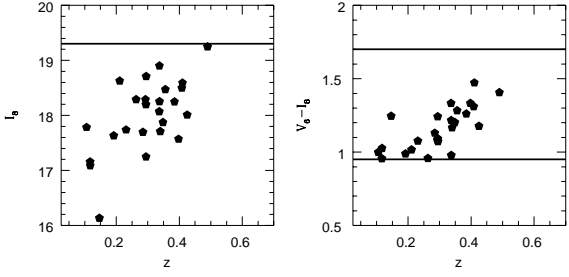}}
\caption{Left panel: observed \Io\, magnitude, as measured by the MDS
group, as a function of redshift. The limit for sample selection
(\Io~$<19.3$) is plotted as horizontal line. Right panel: observed
\Vs-\Io\, colour, as measured by the MDS group, as a function of
redshift. The limits for sample selection ($0.95<$\Vs$-$\Io$<1.7$) are
plotted as horizontal lines.}
\label{fig:colormagz}
\end{figure}

For a physical interpretation of the results it is crucial to have a
full control of the relevant selection effects. In particular, from
the point of view of the evolution of stellar populations it is
important to quantify the effects of the colour selection and the
magnitude limit. As described in PII, we selected our galaxies from
the MDS catalog\footnote{By comparing our photometry with that of the
MDS group we find an average difference of less than 0.01 both in
colour and \Io\, magnitude and an r.m.s. scatter of 0.11 and 0.12
respectively in colour and \Io\, magnitude. The scatter is taken into
account in the treatment of the selection effects.}, according to
\Io~$<19.3$ and $0.95<$~\Vs-\Io~$<1.7$. In Figure~\ref{fig:colormagz}
we plot magnitudes and colours of the galaxies in the observed sample,
together with the selection criteria represented by solid lines.

Two main potential biases can be introduced by our selection. On the
one hand, at high redshift ($z\ga 0.4$), the typical luminosity
approaches that of the selection criterion in \Io. On the other hand,
at low redshift ($z=0.1-0.2$), the colours of the galaxies in our
sample approach the lower limit in the selection criterion. In this
section we will focus mostly on the former potential bias, while the
latter will be better addressed in the next section, with the help of
evolutionary population synthesis models.

Some examples clarify these selection effects. Let us first consider a
toy population of galaxies with the following properties. For given
$\sigma$ and \Rekpc, the scatter in \sbe\, results from measurement
errors and the intrinsic scatter of the FP; the colour is not
correlated with the offset of the intercept. For this toy population,
the magnitude limit will tend to select the brightest objects, thus
simulating a stronger evolution of the intercept of the FP. The
magnitude of the bias depends on the measurement error and the
intrinsic scatter of the FP. This effect can be quantified as
follows. Let us assume that at $z=0.4$ (the highest redshift, where
the effect is stronger) field early-type galaxies define an FP with
slopes $\alpha$ and $\beta$ equal to those of Coma. The built-in
intercept is offset by $\Delta\gamma_{th}$ with respect to the FP of
Coma, and the scatter is kept as a parameter. A population of galaxies
is created by a routine implementing a Montecarlo algorithm, the
magnitude limit \Io~$<19.3$ applied, and the observed offset $\Delta
\gamma_{obs}$ recovered. This is repeated for a range of values of
$\Delta \gamma_{th}$.

\begin{figure}
\mbox{\epsfxsize=8cm \epsfbox{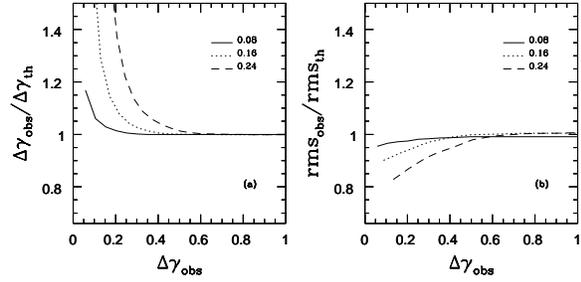}}
\caption{Estimate of the magnitude selection bias. Panel (a): due to
the magnitude limit (see Fig.~\ref{fig:colormagz}) the value of
$\Delta \gamma$ recovered from observation ($\Delta \gamma_{obs}$) can
be higher than the input value ($\Delta \gamma_{th}$); here we plot
the ratio of the observed to input value as a function of observed
value. Panel (b): The magnitude limit can also bias the observed
scatter (rms$_{obs}$); here we plot the ratio of the observed scatter
to the input scatter (rms$_{th}$). The curves are derived with
Montecarlo simulations at $z=0.4$, i.e. the highest redshift bin,
where the effects are stronger.  The curves are labelled with the
intrinsic scatter in $\gamma$ of the FP. The $\Lambda$ cosmology is
assumed. Similar results are obtained for the classical cosmology.}
\label{fig:aplotg0.3}
\end{figure}

In Figure~\ref{fig:aplotg0.3}, panel (a), we plot the ratio $\Delta
\gamma_{obs}/\Delta \gamma_{th}$ as a function of $\Delta \gamma_{obs}$
and for various values of the intrinsic scatter. As expected, the bias
is important for low values of $\Delta \gamma$ and for high values of
the scatter. For our observed values ($\Delta
\gamma_{obs}\approx 0.2-0.25$, scatter $\approx 0.08$) the effect is
small ($\la5$ per cent ). In addition, we can use the same simulation to
check whether the selection effects introduce any bias in the
determination of the scatter. In panel (b) we plot the ratio of the
recovered scatter (rms$_{obs}$) to the input scatter (rms$_{th}$) as a
function of $\Delta \gamma_{obs}$. For $\Delta
\gamma_{obs}\approx0.2-0.25$ the systematic effect is much smaller
($<10$ per cent) than the random error. 

In a more realistic second example, let us consider a population of
early-type galaxies with fixed metallicity and a broad distribution of
ages. At $z\approx 0.4$ our magnitude limit would tend to select the
brightest (i.e.  the youngest and bluest) objects. At $z=0.1-0.2$, the
colour selection criterion will pick the reddest (i.e. the oldest and
faintest) objects. In this case the colour and magnitude selection
effects are connected, and some modeling of the evolution of the
galactic spectra is required to study the bias. The next section
reports a quantitative analysis of these problems based on a
Bayesian-Montecarlo approach.

\section{Constraints on the evolution of stellar populations}

\label{sec:evo}

In this section we analyse the results presented above in terms of
passive evolution of the stellar populations. First, in
\ref{ssec:sbm}, we compare the data to the prediction of single-burst
passive evolution models based on Bruzual \& Charlot (1993; GISSEL96
version, hereafter BC96) synthetic spectra. In this first qualitative
analysis selection biases are neglected.  In the following
Section~\ref{ssec:scMa} we introduce a Bayesian-Montecarlo approach
(see below and the description of the algorithm in Treu 2001), useful
to derive quantitative results, taking into account the selection
process. In its simplest version, the approach is used to calculate
the {\it a posteriori} probability density (i.e. the probability
density given the present set of observations) of the $\tau$ parameter
defined in Equation~\ref{eq:gammev}. Then the Bayesian-Montecarlo
approach is combined with the BC96 spectral synthesis models to derive
constraints on the star formation history of the galaxies in the
sample.

As is well known, the problem of determining the star formation
history of stellar populations from the Fundamental Plane (and the
\sbe-\Rekpc\, relation) is sensitive to the value of the cosmological
parameters and the modeling of the stellar populations. In particular,
it depends on the choice of the Initial Mass Function (IMF; see Schade
et al.\ 1997). For this reason, the evolution of the FP can be used to
constrain the cosmological parameters and/or the IMF (Bender et al.\
1998; van Dokkum et al.\ 1998b).  In the present study, in order to
limit this sort of degeneracies, we focus on the `best-guess' for the
value of the cosmological parameters and the form of the IMF
($\Lambda$ cosmology and Salpeter IMF) and then we try to extract the
maximal information on the star formation history. In order to give an
idea of what is robust and what is sensitive to the assumptions about
the cosmological parameters and the IMF, we will also present the
results for the classical cosmology and for the Scalo IMF. For
completeness we briefly discuss (Section~\ref{ssec:cosmo}) the limits
that can be set on the cosmological parameters.
 
In Section~\ref{ssec:sys} we discuss potential systematic biases and
in Section~\ref{ssec:aeo} we extend the discussion of the
population biases introduced in Section~\ref{sec:FPz}.

\subsection{Single-burst models}

\label{ssec:sbm}

\begin{figure}
\mbox{\epsfysize=8cm \epsfbox{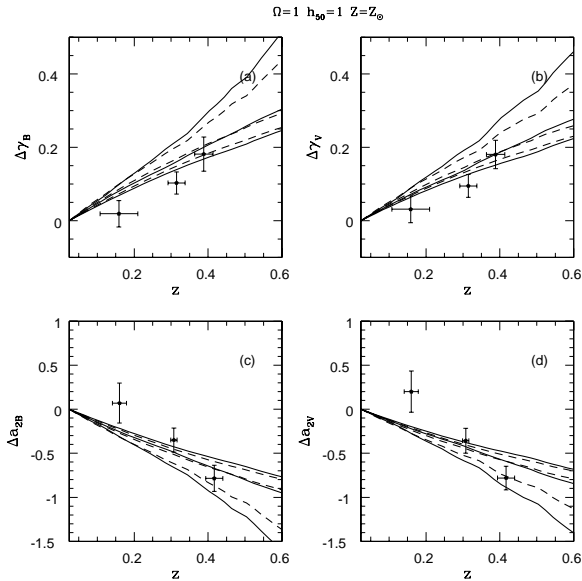}}
\caption{Single-burst passive evolution models. The evolution of the
intercept of the FP is compared to the prediction of single-burst
passive evolution models computed from BC96 synthetic spectra in
panels (a) and (b). The comparison is repeated in panels (c) and (d)
for the \sbe-\Rekpc\, relation, by including galaxies without measured
velocity dispersion. The solid lines represent models with Salpeter
IMF (Salpeter 1955), the dashed lines represent models with Scalo IMF
(Scalo 1986). All models have solar metallicity. Three redshifts of
formation are assumed, $z_f=1,2,5$, from top to bottom in panels (a)
and (b), from bottom to top in panels (c) and (d). The classical
cosmology is assumed.}
\label{fig:evo1solar}
\end{figure}

\begin{figure}
\mbox{\epsfysize=8cm \epsfbox{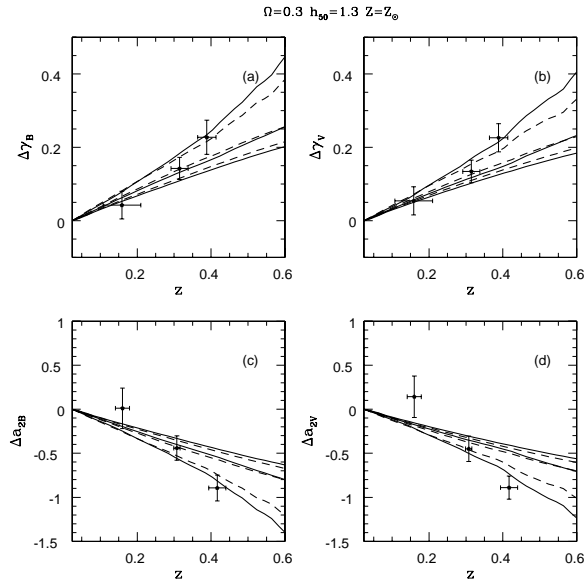}}
\caption{As in Figure~\ref{fig:evo1solar}, in the $\Lambda$ cosmology.}
\label{fig:evo0.3solar}
\end{figure}

In Figures~\ref{fig:evo1solar} and \ref{fig:evo0.3solar} the evolution
of $\gamma$ and $a_2$ as predicted by single-burst passive evolution
models is shown and compared to the data.  For each cosmological model
we show the predictions obtained for single bursts occurred at
$z_f=1,2,5$. The results are shown for models with solar metallicity
and kl96 stellar atmospheres (see the documentation to BC96 for
details). The results are highly insensitive to changes in
metallicities and stellar atmospheres.

Qualitatively, we note that the data points are better reproduced by
models with $z_f\ga2$ in the classical cosmology and with $z_f\sim1$
in the $\Lambda$ cosmology.

\subsection{A Bayesian-Montecarlo approach to the evolution of stellar populations}

\label{ssec:scMa}

We can better quantify the results obtained so far by means of a
Bayesian-Montecarlo approach. In the following we will give only a
brief description of the method, which is outlined in detail elsewhere
Treu (2001).

\subsubsection{The evolution of the intercept of the FP}

Let us assume that Equation~\ref{eq:MLzg} holds and that the intercept
$\gamma$ evolves as in Equation~\ref{eq:gammev}, i.e. linearly in
redshift with slope $\tau$ ($\Delta \gamma= \tau z$). Under these
assumptions we are able to calculate the probability
$p^i=p^i(\gamma^i|\tau)$ of observing a galaxy at redshift $z^i$ with
$\gamma^i$ (defined as in Equation~\ref{eq:gammai}), where the
superscript $i$ indicates an individual galaxy within the set of
observed galaxies. The probability density is obtained from Montecarlo
simulations (at least 1,000,000 random events per galaxy are
generated, see Treu 2001), taking into account the intrinsic scatter
of the FP (assumed to be constant with redshift as measured above),
the observational errors, and the selection effects. Since galaxies
are independent, the probability of observing the set \{$\gamma^i$\}
is:
\begin{equation}
p(\{\gamma^i\}|\tau)=\Pi_i p^i({\gamma^i|\tau}).
\label{eq:ppp}
\end{equation}
We now apply Bayes Theorem (Bayes 1763; see also, e.g., Zellner 1971):
\begin{equation}
p(\tau|\{\gamma^i\}) = \frac{p(\{\gamma^i\}|\tau) p(\tau)} {\int
p(\{\gamma^i\}|\tau) p(\tau) d\tau},
\label{eq:Bayeus}
\end{equation}
to derive the probability of a value of $\tau$ given the set of
measured $\{\gamma^i\}$ (the {\it a posteriori} probability, or simply
the posterior). The probability density of $\tau$ prior to
observations (the {\it a priori} probability, or simply the prior) can
be used to take into account previous observations or physical
limits. The choice of the {\it a priori} probability density in cases
where investigators know little about the value of the parameter, or
wish to proceed as if they knew little, is extensively discussed in
the literature (see, e.~g., Zellner 1971 and references therein; see
also the recent review by Berger 1999). Traditionally, in the
so-called Bayes-Laplace approach, a uniform prior was
used. Alternative priors have been proposed, including some based on
the requirement of invariance with respect to changes of
parametrization (Jeffreys 1961) and others based on information
arguments (see next paragraph; see also Zellner 1971). The advantage
of the Bayesian approach is that it provides a probability
distribution for the parameter (as compared for example to the
confidence limits provided by a 'frequentist' approach), under
well-controlled assumptions (such as the choice of the prior).

For simplicity, in the following we assume $p(\tau)$ constant within
the interval $(-0.1,1.5)$ and zero outside it, i.e.
\begin{equation}
p(\tau)=0.625 H(1.5-\tau) H(\tau+0.1), 
\label{eq:priori}
\end{equation}
with $H$ the Heaviside function, defined as $H(x)=1$ for
$x>0$ and 0 otherwise. The interval has been chosen so as to include
all the region where $p(\{\gamma^i\}|\tau)\ne 0$. It is interesting to
note that for a finite interval spanned by the parameter, the uniform
prior is the proper (i.e. normalizable to 1) prior with minimal
information (defined as $\int p(x)\log p(x) dx$; Zellner 1971). It is
also noticed that for this choice of the prior the maximum of the
posterior is the value that would be derived with a maximum likelihood
algorithm.

\begin{figure}
\mbox{\epsfxsize=8cm \epsfbox{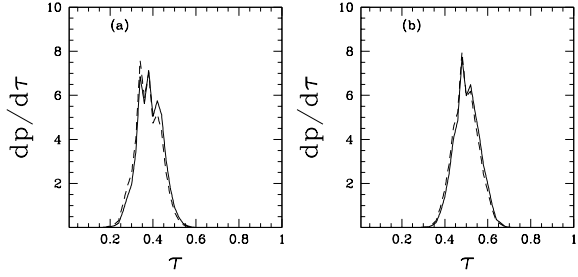}}
\caption{Evolution of the intercept of the FP in the V band. The {\it
a posteriori} probability density of the slope $\tau$ of the relation
$\Delta \gamma = \tau z$ is shown here as recovered from a
Bayesian-Montecarlo approach that takes into account selection biases
(see Section~\ref{ssec:scMa}). The posteriors obtained with a uniform
prior (solid line) and with a prior of the form $p(\tau)\propto
1/\tau$ (dashed line) are shown. The results for the classical
cosmology and the $\Lambda$ cosmology are shown in panels (a) and
(b).}
\label{fig:tau2}
\end{figure}

The method described above, extended to the galaxies without measured
velocity dispersion in terms of the \sbe-\Rekpc\, relation as outlined
by Treu (2001), gives the results shown in Figure~\ref{fig:tau2} (V
band).
One sigma limits on $\tau$ can be defined in the usual way, by
requiring that the integral of the probability density over the
interval be 0.68. In this way, we find $0.33<\tau<0.44$ (classical
cosmology) and $0.44<\tau<0.56$ ($\Lambda$ cosmology). These intervals
are consistent with the values found using a least $\chi^2$ fit (see
Equation~\ref{eq:gammev} and the following discussion). However, the
values found with the least $\chi^2$ fit are on the higher end of the
interval, not on the peak of the probability density (see
Figure~\ref{fig:tau2}). We conclude that a correct treatment of the
selection effects is important in order to derive reliable estimates.

In order to check that the results are robust with respect to the
choice of the prior, we computed the posterior by assuming an
alternative prior. For example, let us consider $p(\tau)\propto1/\tau$
(limited to $\tau>0$), which has the advantage that the results are
unchanged under a power law change of parametrization $X=\tau^n$ (see
Zellner 1971 and references therein). In fact, by converting
\begin{equation}
p(X|\{\gamma^i\})\propto p(\{\gamma^i\}|X) p(X) \propto \frac{p(\{\gamma^i\}|X)}{X}
\end{equation}
into a posterior in $\tau$ we get
\begin{equation}
p'(\tau|\{\gamma^i\}) \propto p(X(\tau)|\{\gamma^i\})
\left|\frac{dX}{d\tau}\right| \propto \frac{p(\{\gamma^i\}|X(\tau))}{\tau},
\end{equation}
as is obtained by assuming $p(\tau)\propto1/\tau$.  The posterior
obtained by taking this prior is shown in Figure~\ref{fig:tau2} as a
dashed line; it is very similar to the one obtained by assuming a
uniform prior (solid line).

\subsubsection{Single-burst models}

The Bayesian-Montecarlo approach is used with the BC96 modeling of
integrated stellar populations to obtain quantitative information on
the star formation history of our sample of field early-type galaxies.
Let us first consider a single population of galaxies with fixed IMF
and stellar populations formed in a single-burst at $z=z_f$ and
evolving passively thereafter. Let us assume that in the local
Universe such population obeys the FP relation of Coma. As usual, we
will assume that the properties of the stellar populations are the
only factor inducing evolution. Therefore, only the intercept of the
FP evolves with redshift, while the scatter and slopes $\alpha,\beta$
remain constant. In this model, for any given redshift, velocity
dispersion, effective radius ($z^i$, $\sigma^i$, $r_{\tx{e}}^i$), and
observational errors of an individual galaxy, we can calculate the
probability density of the colours, magnitude, and intercept of the FP
($\gamma^i$ defined in Equation~\ref{eq:gammai}). By taking into
account the selection effects with a Montecarlo routine (a large
number of points are generated and only the ones satisfying the
selection criteria are kept) we obtain the probability density
$p^i(\gamma^i|z_f)$ of observing $\gamma^i$ given $z_f$.  Since the
galaxies are independent samplings of the statistical ensemble we
obtain:
\begin{equation}
p(\{\gamma^i\}|z_f)=\Pi_i p^i(\gamma^i|z_f).
\end{equation}
As in Equation~\ref{eq:Bayeus}, the probability density of $z_f$ given
the set of observations can be computed by assuming an {\it a priori}
probability density $p(z_f)$ from the Bayes theorem:
\begin{equation}
p(z_f|\{\gamma^i\}) = \frac{p(\{\gamma^i\}|z_f) p(z_f)} {\int
p(\{\gamma^i\}|z_f) p(z_f) dz_f}.
\label{eq:Bayeuszf}
\end{equation}
In the following, we will assume a constant {\it a priori} probability
between two values of $z$:
\begin{equation}
p(z_f)=p_1 H(z_f-z_m)H(z_M-z_f).
\label{eq:apriorizf}
\end{equation}
A very natural choice is one with $z_m$ very close to the highest
$z^i$ of the sample and a high value of $z_M$, in order to include the
interval where $p(\gamma^i|z_f)$ is non-zero. We use $z_m=0.5$ and
$z_M=4.0$ for the $\Lambda$ cosmology and $z_m=0.5$ and $z_M=10.5$ for
the classical cosmology. 

\begin{figure}
\mbox{\epsfysize=8cm \epsfbox{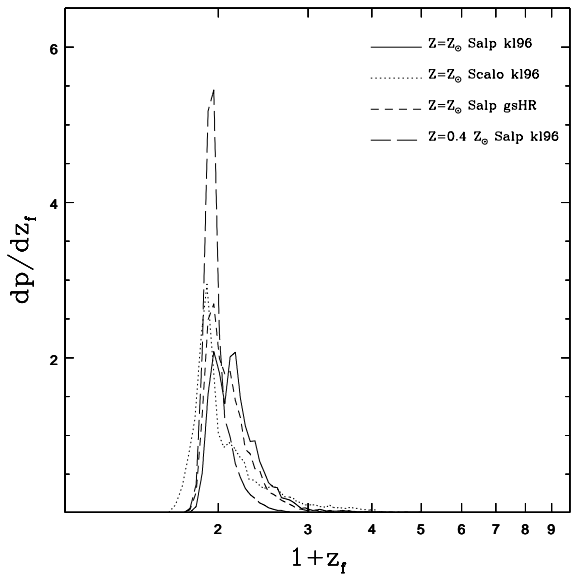}}
\caption{Single-burst stellar population models, I. The probability
density of the redshift of formation ($z_f$) of the stellar
populations of field early-type galaxies, given the present set of
observations, is shown for various realizations of the models. BC96
spectral synthesis models are used with Salpeter or Scalo IMF, solar
or 0.4 solar metallicity, stellar atmospheres from kl96 or from gsHR
(see Bruzual \& Charlot 1996 for details). Independently of the model,
the probability density turns out to peak at $z_f\sim 1$. The
$\Lambda$ cosmology is assumed. Details of the method are given in
Section~\ref{ssec:scMa}.}
\label{fig:zfo0.3}
\end{figure}

\begin{figure}
\mbox{\epsfysize=8cm \epsfbox{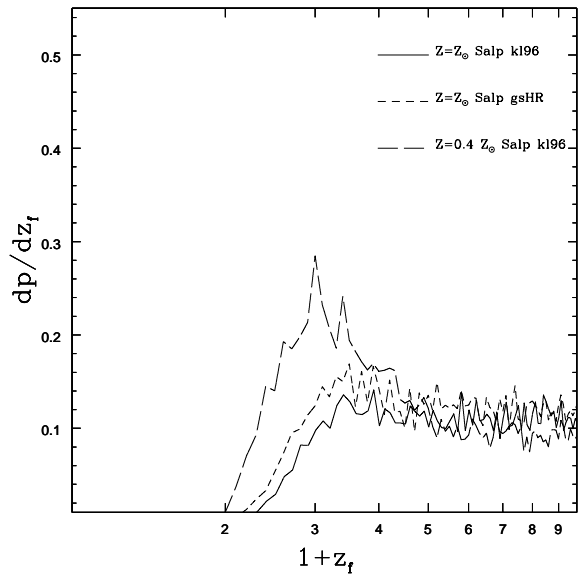}}
\caption{Single-burst stellar population models, II. 
As in Figure~\ref{fig:zfo0.3}, for the classical cosmology.}
\label{fig:zfo1}
\end{figure}

The {\it a posteriori} probability densities are shown in
Figures~\ref{fig:zfo0.3} and \ref{fig:zfo1}. The probability density
peaks at $z\approx 1-1.2 $ (8-9 Gyr look-back time) and it extends to
$z\approx 0.8$ (7 Gyr) and $z\approx1.8$ (11 Gyr) in the $\Lambda$
cosmology. In the classical cosmology the probability is much flatter
and it extends from $z\approx 1.5-2$ (10 Gyr) to the upper limit of
the sampled interval (the age of the Universe in this cosmology is
13.1 Gyr). The distributions are computed for a variety of
realizations of the model, with different metallicities, IMFs, stellar
atmospheres, showing that the result is robust and does not depend on
the details of the model.

In order to check the dependence of the results on the {\it a priori}
probability density we consider the following approach. An
alternative and equivalently straightforward {\it a priori}
probability is the one obtained by considering the look-back time of
formation ($t_f$) as the parameter of the model, instead of the
redshift of formation ($z_f$) used in Equation~\ref{eq:Bayeuszf}:
\begin{equation}
p(t_f|\{\gamma^i\}) = \frac{p(\{\gamma^i\}|t_f) p(t_f)} {\int
p(\{\gamma^i\}|t_f) p(t_f) dt_f}.
\label{eq:Bayeustf}
\end{equation}
Similarly to the case with $z_f$, we assume a prior uniform within the
interval defined by the look-back time of the most distant galaxy in
our sample and the age of the Universe.  The posterior
$p(t_f|\{\gamma^i\})$ can be transformed into the posterior in $z_f$
in the usual way:
\begin{equation}
p'(z_f|\{\gamma^i\})=p(t_f(z_f)|\{\gamma^i\})\left|\frac{dt_f}{dz_f}\right|,
\label{eq:transprob}
\end{equation}
where
\begin{equation}
\left|\frac{dt_f}{dz_f}\right|=\frac{1}{H_0(1+z_f)\sqrt{\Omega(1+z_f)^3+
\Omega_R(1+z_f)^2+\Omega_{\Lambda}}},
\label{eq:dtdz}
\end{equation}
with
\begin{equation}
\Omega_R=1-\Omega_{\Lambda}-\Omega.
\end{equation}

Since $p(\{\gamma^i\}|z_f) = p(\{\gamma^i\}|t_f(z_f))$, from
Equations~\ref{eq:Bayeuszf},~\ref{eq:Bayeustf}, and~\ref{eq:transprob}
it follows that:
\begin{equation}
p'(z_f|\{\gamma^i\})\propto p(z_f|\{\gamma^i\})\left|\frac{dt_f}{dz_f}\right|,
\label{eq:pdtpdz}
\end{equation}
i.e. the probability density obtained by assuming a prior uniform in
$t_f$ is proportional to the one obtained by assuming a prior uniform
in $z_f$ multiplied by the modulus of the derivative of $t_f$ with
respect to $z_f$ (the constants are set by requiring that the
probability density integrated over $z_f$ be equal to 1).

\begin{figure}
\mbox{\epsfysize=8cm \epsfbox{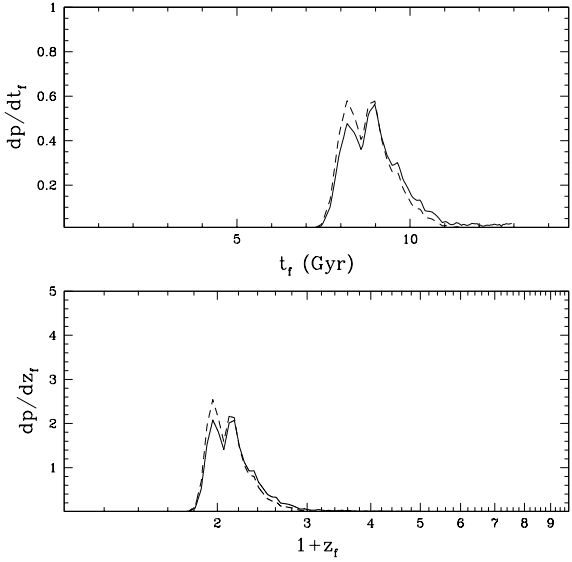}}
\caption{Single-burst stellar population models, III. 
Effects of the {\it a priori} probability density. The {\it a posteriori} probability
density of the look-back time of formation ($t_f$) and of the redshift
of formation ($z_f$) are plotted respectively in the upper and lower
panel ($\Lambda$ cosmology). Dashed lines correspond to a prior
uniform in $t_f$, solid lines to a prior uniform in $z_f$. BC96
spectral synthesis models are used with Salpeter IMF, solar
metallicity, and kl96 stellar atmospheres. The effects
of changing the {\it a priori} probability are negligible. A discussion of the
{\it a priori} probability density is given in Section~\ref{ssec:scMa}.}
\label{fig:tzfo0.3}
\end{figure}

\begin{figure}
\mbox{\epsfysize=8cm \epsfbox{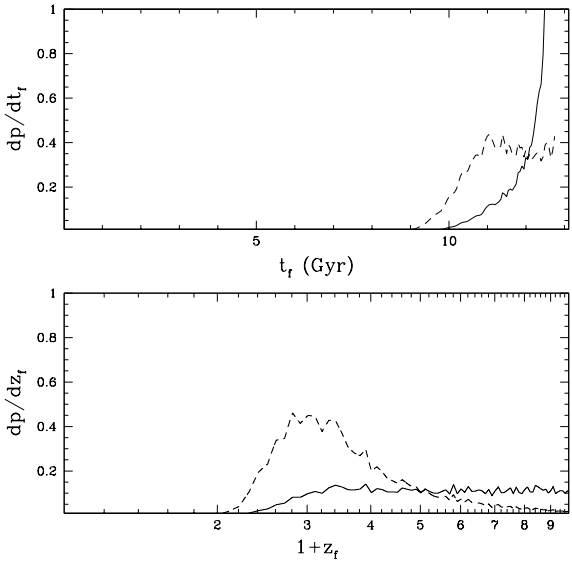}}
\caption{Single-burst stellar population models, IV. As in 
Figure~\ref{fig:tzfo0.3}, for the classical cosmology (see
Section~\ref{ssec:scMa}).}
\label{fig:tzfo1}
\end{figure}

The resulting {\it a posteriori} probability densities are plotted in
Figures~\ref{fig:tzfo0.3} and~\ref{fig:tzfo1}, respectively for the
$\Lambda$ and the classical cosmology.
It is noticed that the {\it a posteriori} probability density changes
negligibly in the $\Lambda$ cosmology, while in the classical
cosmology the probability density peaks at smaller $z_f$ and declines
at high $z_f$. However, this does not alter the physical
interpretation of the result, since in any case the stellar
populations turn out to be formed long ago ($\ga 10$ Gyr). Since
the effects of changing the {\it a priori} probability density do not alter the
physical interpretation of the results, in the following we will
present only the results obtained by assuming a prior uniform in
$z_f$, recalling that the results obtained by assuming a prior uniform
in $t_f$ can be obtained with Equation~\ref{eq:pdtpdz}.
\subsubsection{Secondary bursts of star formation}

The single-burst stellar population model is useful as a benchmark to
quantify and compare different results. However, it is clearly a
simplified picture of the star formation history of early-type
galaxies. A small amount ($<10$ per cent) of the total mass of moderately
young stars (1 Gyr or so) can significantly alter the integrated
colours and mass-to-light ratio of an old stellar population for a few
Gyr. In this scenario, after a few Gyr, the integrated colours and
mass-to-light ratio become totally indistinguishable from the ones of
an old single-burst stellar population (see also Jimenez et al.\
1999).

\begin{figure}
\mbox{\epsfysize=8cm \epsfbox{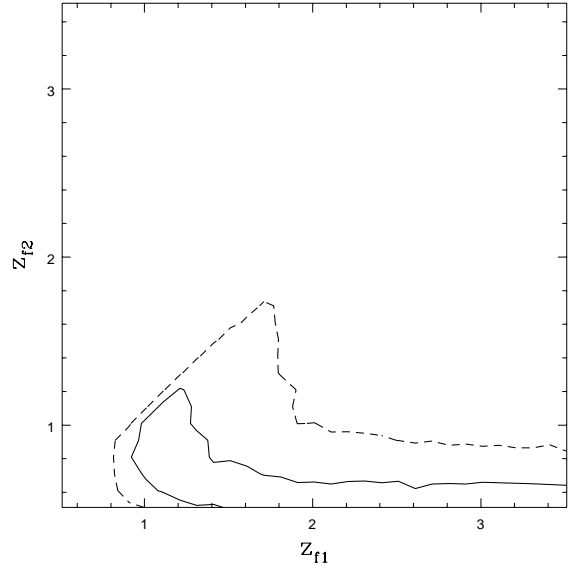}}
\caption{Model with two populations of stars. The older component is formed at
$z_{f1}$, the younger, with a tenth of the mass, is formed at
$z_{f2}$.  Contour levels of the {\it a posteriori} probability are shown as
solid (68 per cent) and dashed (95 per cent) lines. The $\Lambda$ cosmology is
assumed. BC96 models with solar metallicity, Salpeter IMF and kl96
atmospheres are used.}
\label{fig:2Dzf}
\end{figure}

A wealth of observations suggest that minor episodes of star formation
can occur from intermediate redshifts to the present (e.g. Schade et
al.\ 1999; Trager et al.\ 2000a; Bernardi et al.\ 1998). Therefore, it
is worth investigating this possibility for our sample of field
early-type galaxies. The task can be accomplished by modifying the
model described above to allow for a secondary burst of star formation
occurred at $z_{f2}<z_{f1}$, being $z_{f1}$ associated with the epoch
of the first burst. In this case Equation~\ref{eq:Bayeuszf} is changed
to
\begin{equation}
p(z_{f1},z_{f2}|\{\gamma^i\}) = \frac{p(\{\gamma^i\}|z_{f1},z_{f2}) p(z_{f1},z_{f2})} {\int
p(\{\gamma^i\}|z_{f1},z_{f2}) p(z_{f1},z_{f2}) dz_{f1}dz_{f2}}.
\label{eq:Bayeuszf2D}
\end{equation}
The {\it a priori} probability is assumed to be constant within a test interval
\begin{eqnarray}
p(z_{f1},z_{f2}) \propto & H(z_{f1}-z_m)H(z_M-z_{f1})\times\nonumber\\
&H(z_{f2}-z_m)H(z_M-z_{f2})H(z_{f1}-z_{f2}),
\label{eq:apriorizf2D}
\end{eqnarray}
similarly to the one-dimensional case. The contour levels of the
probability density corresponding to 68 per cent and 95 per cent probability are
shown in Figure~\ref{fig:2Dzf} for the $\Lambda$ cosmology and a mass
ratio of the two populations of 10 (the older being 10 times the mass
of the younger). Not surprisingly, it is sufficient to have a small
mass of stars formed at $z<0.6-0.8$ to make it possible for the rest
of the stellar mass to be formed at the very beginning of the Universe
($z_{f1}\sim 3-4$, i.e. 12-13 Gyr ago).

It is important to check that such scenario is also consistent with
the small scatter of the FP observed at low and intermediate redshift.
Since a number of physical phenomena contributes to the scatter of the
FP, the scatter due to differences in ages must be smaller than the
observed one.  In order to simulate this effect we have considered a
very simple model, where the bulk of the stellar population is formed
at $z=z_b$ and a secondary burst (of a tenth of the mass) occurs at a
later time, with uniform probability between $z=z_{s1}$ and
$z=z_{s2}$. The observed trend in $\Delta\gamma$ vs. $z$ is reproduced
without any particular fine tuning of the parameters ($z_b\sim
2-3$, $z_{s1}\approx0.6$, $z_{s2}\approx0.8$, for example, work well)
and the resulting scatter in $\gamma$ is very small ($\la 0.01$).  It
is interesting to note that, if we extend to lower redshifts the tail
of the secondary burst of star formation, the evolution of $\gamma$ is
not altered significantly, while the scatter induced by the spread in
average age of the stellar populations is comparable to the observed
one ($\la 0.08$).

\subsection{Constraining the cosmological parameters}

\label{ssec:cosmo}

\begin{figure}
\mbox{\epsfysize=8cm \epsfbox{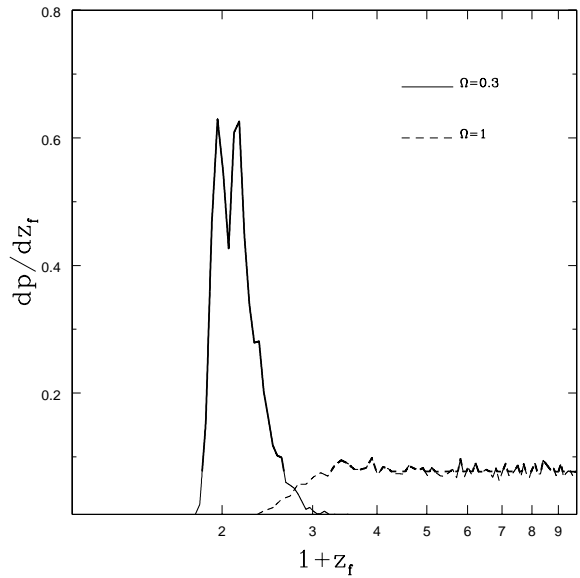}}
\caption{Probability density of the redshift of star-formation given the
present set of observations. The {\it a posteriori} probability densities for
the classical and the $\Lambda$ cosmology are compared by means of an
extension of the Bayesian-Montecarlo approach (see
Section~\ref{ssec:cosmo}). The maximum of the probability is for the
$\Lambda$ cosmology. However, the 68~per cent limits (thick lines) include
also the classical cosmology.}
\label{fig:cosmo}
\end{figure}

The Bayesian-Montecarlo approach can be extended also to constrain the
cosmological parameters. To this aim, it is sufficient to modify
Equation~\ref{eq:Bayeuszf} into
\begin{equation}
p(z_f,\Omega |\{\gamma^i\}) = \frac{p(\{\gamma^i\}|z_f,\Omega) p(z_f,\Omega)} {\int
p(\{\gamma^i\}|z_f,\Omega) p(z_f,\Omega) dz_f d\Omega},
\label{eq:Bayeuszfomega}
\end{equation}
where, for simplicity, only $\Omega$ has been introduced, and to
assume a discrete {\it a priori} probability using the Dirac $\delta$
distribution
\begin{eqnarray}
p(z_f,\Omega)= & \bar{p}_1 H(z_f-z_{m1})H(z_{M1}-z_f)\delta(\Omega-0.3)\nonumber\\&+ \bar{p}_2 H(z_f-z_{m2})H(z_{M2}-z_f)\delta(\Omega-1).
\label{eq:apriorizfomega}
\end{eqnarray}
The normalization factors ($\bar{p}_1,\bar{p}_2$) are chosen so that
the integral of each term over $z_f$ and $\Omega$ equals 1/2. The
intervals in $z_f$ are chosen as in Section~\ref{ssec:scMa}. The
probability density $p(\{\gamma^i\}|z_f,\Omega )$, is related to the
probability densities in Equation~\ref{eq:Bayeuszf} by
\begin{eqnarray}
p(\{\gamma^i\} | z_f, \Omega ) \propto & p_{\Omega =0.3}(\{\gamma^i\}|z_f)\delta(\Omega - 0.3) \nonumber\\ & + p_{\Omega =1}(\{\gamma^i\}|z_f) \delta (\Omega - 1).
\end{eqnarray}
The probabilities obtained for the two cosmologies are compared in
Figure~\ref{fig:cosmo}. The maximum of probability occurs clearly for
the $\Lambda$ cosmology but the area subtended by the classical
cosmology is actually larger (0.3 vs 0.7). Hence, at this redshift,
given the present uncertainties on the evolution of the stellar
populations, the $\Lambda$ cosmology is the `best-guess' cosmology,
but the classical cosmology cannot be ruled out by this method (see
the 68 per cent limits marked with thick lines in
Figure~\ref{fig:cosmo}; see also the discussion by Bender et al.\ 1998
and van Dokkum et al.\ 1998b).

\subsection{Systematic errors}

\label{ssec:sys}

The analysis performed above takes the random errors of the
measurement into account. However, we still have to estimate the
uncertainty related to global systematic errors, such as:

\begin{enumerate}
\item {\bf Zero point errors}. We estimate the uncertainty in the absolute
zero point of filters F606W and F814W to be $\la 0.03$ mag (see the
HST Instrument Handbook and references therein for the listing and
description of uncertainties)
\item {\bf Errors in the K-color correction}. These errors are $\sim
0.01-0.02$ mag (see PII). As a check, we compared the K-color
corrections obtained with our method with the ones derived by Kelson
et al.\ (2000a) and van Dokkum et al.\ (1998b). We found, consistent
with the error estimate, differences $\sim 0.01-0.02$ mag.
\item {\bf Uncertainty in the local FP}.  The slopes and the zero point of
the FP of the Coma cluster are known with errors. The uncertainty on
$\gamma$ for our reference samples (Bender 1998, Lucey 1991) is $\sim
0.01$, including photometric calibration, K-correction, and foreground
extinction. We assume the distance to Coma to be $cz=7200$\,\kms\, and
we neglect, as discussed in Section~\ref{sec:obs}, differences between
the FP of Coma and the FP of other clusters or the field. We also
neglect possible small differences between the FP in the centre and in
the outer parts of Coma (Lucey et al.\ 1991).
\end{enumerate}

The total systematic error, obtained by adding in quadrature the terms
listed above, does not exceed 0.015 in $\gamma$. We tested the
robustness of our results with respect to this uncertainty by
recomputing the probability density for $\tau$ after changing the Coma
intercept by $\pm0.015$. The 68~per cent limits are changed by $\pm
10$~per cent.

Since the coefficients of the FP and \sbe-\Rekpc\, relation are
measured independently, the very good agreement of the results found
in B and V, with the FP and \sbe-\Rekpc\, relation, is an important
check on the accuracy of the photometric calibration and the
determination of the local relations.

In principle, it is possible to take into account systematic errors in
the Bayesian-Montecarlo approach. In fact, it is sufficient to treat
the photometric zero point, the K-color corrections, and the intercept
of the local FP as additional parameters of the model and then
integrate over the values of the parameters, as in the following
example. Let us denote the set of additional parameters related to the
systematic errors as $\{Y^i\}$, and let us assume that their
probability distribution are independent
$p(\{Y^i\})=\Pi_ip(Y^i)$. Thus, Equation~\ref{eq:Bayeus} becomes:
\begin{equation}
p(\tau|\{\gamma^i\}) = \frac{\int p(\{\gamma^i\}|\tau\{Y^i\})
p(\tau)\Pi_i p^i(Y^i)dY^i} {\int p(\{\gamma^i\}|\tau \{Y^i\}) p(\tau)
\Pi_ip^i(Y^i)dY^i d\tau}.
\label{eq:BayeusY}
\end{equation}
However, the computational cost of implementing this procedure in our
Bayesian-Montecarlo code\footnote{In practice the CPU time (a few hours
on a Pentium III running at 500 Mhz for the computations shown in
Figures~\ref{fig:zfo0.3} to~\ref{fig:tzfo0.3}) is multiplied by the
number of samplings of the distribution of the parameters related to
systematic errors. Even by taking only ten samplings per systematic
parameter ($Y^i$), the CPU time increases 1000 times.} is not
justified in our opinion, given the limited knowledge of the
systematics, and the other sources of uncertainties that enter the
interpretation of the results (see also the next subsection).

\subsection{Morphological evolution bias}

\label{ssec:aeo}

The morphological selection criterion can introduce a bias in our
results (the morphological evolution bias described in
Section~\ref{sec:FPz}). For example, one may think (Franx \& van Dokkum
1996) that early-type galaxies become morphologically recognizable
only after their stellar populations reach a minimum age ($A_0$). In
this case, the evolution of the morphologically selected sample would
be biased towards an older average age of the stellar population
(simply because objects younger than $A_0$ are discarded).

A simple model where E/S0 form at a random time in a given redshift
interval can help quantify the bias. For example, in the $\Lambda$
Universe, by forming galaxies in the redshift range $z=0.5-1.5$, and
by considering them as E/S0 only after 3 Gyr, we can reproduce the
observed evolution of $\gamma$ with constant scatter $\sim 0.06-0.07$.

However, in this scenario no E/S0 would be recognizable at $z=1$, in
contrast with the observations (Im et al.\ 1996; Schade et al.\ 1999;
Treu \& Stiavelli 1999). By requiring that at least one third of
present-day E/S0 be already assembled and recognizable at $z=1$, we
may need a smaller $A_0$ or an earlier period of formation. Reducing
$A_0$ does not work, since the age component in the scatter of the FP
rapidly exceeds the values allowed by the observations. An earlier
period of formation is therefore needed. Starting earlier the
formation ($z\sim3$ is required) alone does not work either, since the
evolution of $\gamma$ becomes too slow. In order not to exceed the
observed scatter it is necessary to stop forming E/S0 earlier in the
Universe. For example, if the formation occurs between $z=1$ and $z=2$
and $A_0=2$ Gyr, the evolution of $\gamma$ is reproduced, the age
contribution to the scatter is always smaller than the scatter
observed, and 48 per cent of E/S0 are recognizable at $z=1$.

In conclusion, based on the results of the FP alone, we cannot exclude
that field E/S0 have been forming until recently ($z=0.5$) if
they become morphologically recognizable only at a moderately old age
($\sim$ 3 Gyr). In turn, we can exclude this possibility, if we take
into account the observational constraint that a significant part of
the population of E/S0 is recognizable at $z=1$. With this constraint
we are forced to push back the epoch of formation of the stellar
populations of field E/S0 to $z\ga1$.

\section{Comparison between field and cluster results}

\label{sec:clu}

In this section we compare the intermediate redshift field FP to
a compilation of cluster data taken from the literature (see
Table~\ref{tab:cluster}).

\begin{table}
\caption{Compilation of sources for FP parameters of galaxies in
clusters at intermediate redshift. For each cluster we list the
redshift, the rest frame band in which photometry is given, the number
of objects per cluster, and the reference. Only objects classified as
E/S0 have been considered. See Section~\ref{sec:clu} for discussion.}
\label{tab:cluster}
\begin{tabular}{lcccl}
Cluster	& z	& band	& nobj 	& ref\\
\hline
CL1358	& 0.33	& V	& 30	& Kelson et al.\ 2000b\\
A370	& 0.375	& B	& 7	& Bender et al.\ 1998\\
MS1512	& 0.375	& B	& 2	& Bender et al.\ 1998\\
CL0024	& 0.39	& V	& 7	& van Dokkum \& Franx 1996\\
MS2053	& 0.58	& V	& 5	& Kelson et al.\ 1997\\ 
MS1054	& 0.83	& B	& 6	& vand Dokkum et al.\ 1998b\\
\hline
\end{tabular}
\end{table}

\begin{figure}
\mbox{\epsfysize=8cm \epsfbox{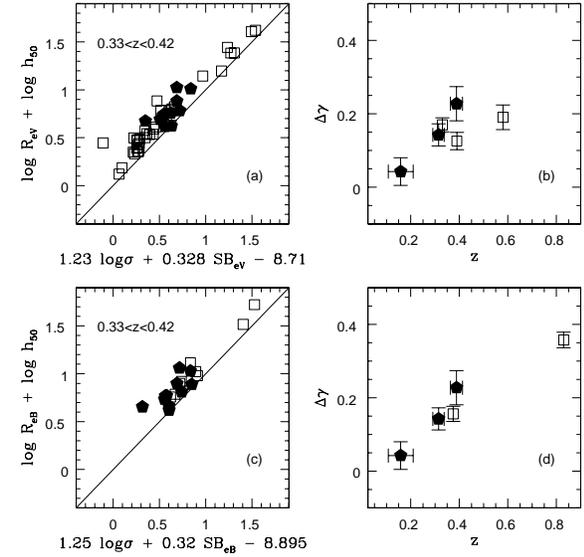}}
\caption{Comparison of field and cluster ellipticals in the FP-space
at intermediate redshift. Filled pentagons are field data points from
this paper, open squares are cluster data taken from the literature
(see Table~\ref{tab:cluster}). In panels (a) and (c) the position in
the FP-space of galaxies at similar redshift is shown. In panels (b)
and (d) the offset of the intercept with respect to the value of Coma
is plotted as a function of redshift. The $\Lambda$ cosmology is
assumed. The comparison yields very similar results in the classical
cosmology (see Treu 2001).}
\label{fig:cfrcluo0.3}
\end{figure}

In Figure~\ref{fig:cfrcluo0.3} we plot the cluster and the field FP
data points in the $\Lambda$ cosmology (the results of the comparison
are very similar for the ``classical'' cosmology and are not shown
here, see Treu 2001). In panels (a) and (c) the cluster and field data
points at $0.33<z<0.42$, respectively in the V and B band, are plotted
in the FP-space. Panels (b) and (d) illustrate the offset of the
intercept with respect to Coma as a function of redshift. The same
slopes $\alpha$ and $\beta$ have been used for consistency, thus
eliminating this source of systematic error in the relative
comparison.
The data at similar redshift ($z\approx 0.3-0.4$) provide the
following results. The clusters points at $z=0.375$ and $z=0.39$
(panels b and d) are marginally (1-2 standard deviations) below the
field data points, while the point at $z=0.33$ is well consistent with
the field data points. By applying the Bayesian-Montecarlo code to
recover the {\it a posteriori} probability density for the cluster
data points, we confirm that: the stellar populations of the galaxies
in the two clusters at $z=0.375$ and $z=0.39$ are marginally older
than the ones in the field. On the other hand, the probability
distribution obtained for the galaxies in the cluster at $z=0.33$ is
indistinguishable from the one obtained in the field. We conclude
that, given the differences in the selection process (see below), and
the size of observational errors, no difference is detected between
cluster and field galaxies.

\begin{figure}
\mbox{\epsfysize=8cm \epsfbox{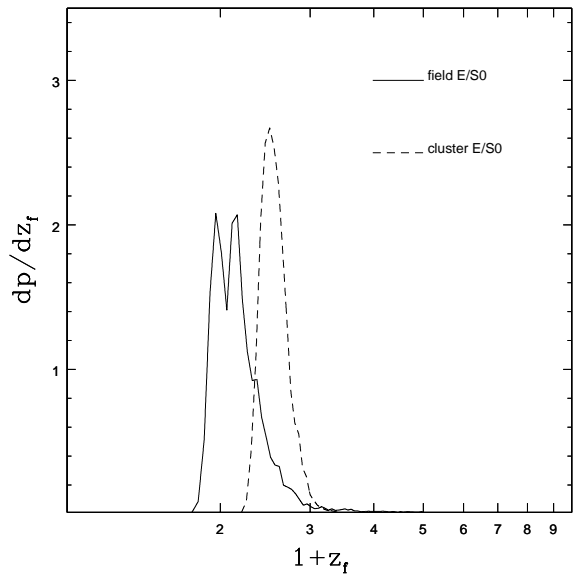}}
\caption{Comparison of the redshift of formation probability density
obtained for field (solid line) and cluster E/S0 (dashed line). The
cluster data in the B band have been considered (see
Table~\ref{tab:cluster}). BC96 models with Salpeter IMF, solar
metallicity, and kl96 atmospheres have been used.  The
$\Lambda$ cosmology is assumed.}
\label{fig:cfrB}
\end{figure}

The cluster data points at higher redshift ($z=0.58-0.83$) show
consistently a smaller offset from the local relation than the one
that would be extrapolated from the field points. For example, if we
apply our Bayesian-Montecarlo approach to the data in B band out to
$z=0.83$, we obtain an older average age for the cluster stellar
populations (see Figure~\ref{fig:cfrB}). A larger sample of field
data, extended to higher redshift, is needed in order to perform a
more stringent comparison. To this aim, we are currently collecting
spectra for 25 galaxies in the redshift range $z=0.4-0.7$ using FORS2
on the Very Large Telescope (VLT). We expect that the observational
errors on $\gamma^i$ will be comparable with the ones obtained for the
present study. Thus, we will be sensitive to possible differences of
$\sim 10$ per cent between the offset of the intercept $\gamma$ for
field and cluster E/S0. 
\subsection{Sample selection effects}

\label{sec:aeoclu}

The galaxies in the cluster sample span wider ranges in $\sigma$ and
\Rekpc\, than those of our sample of galaxies. This effect is related
to the lower limit in velocity dispersion measurable by our
instrumental setup (see PII) and by the absence of giant objects such
as brightest cluster galaxies in our sample. Note that we were not
biased against very large (and hence bright) objects. We simply did
not find any of them.  This selection effect could bias the results
if, for example, larger galaxies were older than smaller ones.

\begin{figure}
\mbox{\epsfysize=8cm \epsfbox{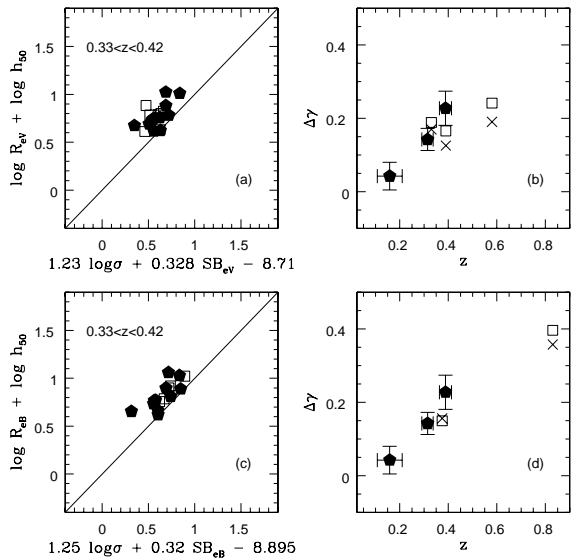}}
\caption{As in Figure~\ref{fig:cfrcluo0.3}. Only cluster galaxies in
the same ranges of $\sigma$ and \Rekpc\, as those of the field sample
have been used (open squares). The results found with the complete
cluster sample are plotted as crosses for reference.}
\label{fig:cfrclus0.3}
\end{figure}

The size of this bias can be estimated by comparing the field data to
a subsample of cluster E/S0 with $\sigma$ and \Rekpc\, within the
limits of our sample. This subsample shows a marginally larger offset
from the local relation than the complete sample (see
Figure~\ref{fig:cfrclus0.3}).
However, this result should be taken with great care, since the
samples we are considering are very small and the change in the
intercepts can be simply due to noise. Larger samples with well
controlled selection effects are needed to address the issue whether
the evolution of the stellar populations depend on effective mass,
possibly with the help of a measurement of a change in the slopes of
the FP with redshift.

In addition, the possible bias induced by the cluster population
evolution (see Section~\ref{sec:FP}) has to be considered. Phenomena
such as the Butcher-Oemler effect (Butcher \& Oemler 1984), the
evolution of the morphology-density relation (Dressler et al.\ 1997),
or the larger scatter of the colour-magnitude relation of lenticular
galaxies, noticed by van Dokkum et al.\ (1998a) in the outer parts of
a cluster at $z=0.33$, seem to suggest that the galaxy population of
clusters evolves. Therefore, it is difficult to extend the information
gathered by studying the FP in the core of clusters at high redshift
to the whole population of present-day clusters. In particular, if
galaxies are accreted from the field into the cluster, the cluster
population at intermediate redshift is not the progenitor of the
present-day cluster population.  Studies with a large field of view,
encompassing the entire range of densities from the core of rich
clusters to the sparse random fields, are needed before we can clarify
how cluster populations evolve and how the environment affects galaxy
evolution.

\section{Conclusions}

\label{sec:consum}

In this paper we have studied the Fundamental Plane and the \sbe-\Rekpc\,
relation for a sample of field early-type galaxies at intermediate
redshift. In particular, we have used the FP and \sbe-\Rekpc\, as
diagnostics of the stellar populations. The main results are:
\begin{enumerate} 
\item {\bf Evolution of the FP}. The FP of field early-type galaxies
exists and is tight out to $z\approx0.4$. The scatter is consistent
with being unchanged out to $z\approx0.4$. The intercept is offset
with respect to that of Coma, in the sense that at a given effective
radius and velocity dispersion galaxies are brighter at $z\approx0.4$
than in the local Universe. The ranges of $\sigma$, \sbe, and
$r_{\tx{e}}$ covered by our sample are not sufficient to measure the
slopes of the FP at intermediate redshift accurately. Similar results
are obtained by studying the \sbe-\Rekpc\, relation out to
$z\approx0.5$.
\item {\bf Single-burst stellar populations}. In a single-burst
scenario, the observed properties of this sample of galaxies are
consistent with those of a single-burst stellar population formed at
$z\approx0.8-1.6$ ($\Lambda$ cosmology) or at $z\ga 2$ (classical
cosmology). These redshifts of formation correspond to a present age
of 7-11 Gyr or 10-13 Gyr, respectively, in the two cosmologies.
\item {\bf Multiple-burst stellar populations}. If a small fraction of 
the stellar mass is formed in a secondary burst at a later time, the
primary burst may have occurred at high $z$ even in the $\Lambda$
cosmology. For example, the data are consistent with the scenario
where the primary burst occurred at $z\ga3$ and a secondary burst,
with a tenth of stellar mass, occurred at $z\approx 0.6-0.8$.
\item {\bf Field vs. Cluster}. No significant difference is found
between the Fundamental Plane of field E/S0 at intermediate redshift
and that of cluster E/S0 at similar redshift taken from the
literature.  The ages of the stellar populations of field early-type
galaxies inferred from the single-burst model are marginally smaller
than the ages derived for samples of cluster early-type galaxies
extended to higher redshift. Unfortunately, various selection
processes, the small samples available, and the different ranges in
global properties spanned by the field and cluster samples do not
allow us to draw firm conclusions. Larger sets of data at higher
redshift, as the one we are currently collecting with the VLT, are
needed to investigate possible differences with higher sensitivity. A
larger set of data, sampling a larger range of effective masses and a
variety of environments from rich clusters to the field, is also
needed to address the issue of the cluster population evolution bias
and the possible dependence of the evolution on effective mass.
\end{enumerate}

\section{Acknowledgments}

This work is based on observations collected at the European Southern
Observatory (La Silla) under programmes 62.O-0592, 63.O-0468, and
64.O-0281 and with the NASA/ESA Hubble Space Telescope, obtained at
the Space Telescope Science Institute, which is operated by
Association of Universities for Research in Astronomy, Inc.\ (AURA),
under NASA contract NAS5-26555. Tommaso Treu was financially supported
by the Space Telescope Science Institute Director Discretionary
Research Fund grant 82228, and by the Italian Ministero
dell'Universit\`a e della Ricerca Scientifica e Tecnologica. Giovanni
Punzi is thanked for useful discussions on Bayesian Statistics.

\end{document}